\documentclass[twocolumn,trackchanges,twocolappendix,tighten]{aastex63}

\usepackage{amssymb, amsmath}
\def\vb#1{\mbox{\boldmath $#1$}}

\received{????}
\revised{????}
\accepted{????}

\submitjournal{ApJ}

\shorttitle{Amplification of turbulence}
\shortauthors{Higashi, Susa \& Chiaki}
\graphicspath{{./}{figures/}}

\begin{document}
\title{Amplification of turbulence in contracting prestellar cores in primordial minihalos}

\email{E-mail:shigashi.phy@gmail.com}

\author[0000-0003-1029-7592]{Sho Higashi}
\affiliation{Department of Physics, Konan University, Okamoto, Kobe, Japan}
\author[0000-0002-3380-5302]{Hajime Susa}
\affiliation{Department of Physics, Konan University, Okamoto, Kobe, Japan}
\author[0000-0001-6246-2866]{Gen Chiaki}
\affiliation{Center for Relativistic Astrophysics, School of Physics, Georgia Institute of Technology, Atlanta, GA, USA}

\begin{abstract}

We investigate the amplification of turbulence through gravitational contraction of the primordial gas in minihalos.
We perform numerical simulations to follow the cloud collapse, assuming polytropic equations of state for different initial turbulent Mach numbers and resolutions.
We find that the turbulent velocity is amplified solely by gravitational contraction, and eventually becomes comparable to the sound speed,  even for small initial turbulent Mach numbers (${\cal M}_0 \gtrsim 0.05$).
We derive an analytic formula for the amplification of turbulent velocity in a collapsing cloud, and find that our numerical results are consistent with the formula.
These results suggest that the turbulence can play an important role in collapsing clouds for general cases.
\end{abstract}


\section{Introduction}\label{sec:intro}
Cosmological simulations based on the {\rm $\Lambda$}CDM model predict that minihalos with $M \sim 10^6~M_{\odot}$ host the first generation of stars in the universe at redshifts $z \gtrsim 10$ (\citealt{Haiman96,Tegmark97, Nishi99,Fuller00,Abel02,Bromm02,Yoshida03}).
It has been believed that the first stars are largely massive ($\sim 10$--$1000~M_{\odot}$) due to inefficient cooling in the primordial gas clouds \citep{Omukai98,Yoshida08}.
However, recent numerical studies have revealed that the fragmentation of accretion disks can lead less massive stars ($\sim 1~M_{\odot}$) to form \citep{Clark11Sci,Greif11,Greif2012,Susa13,Susa14,Stacy16,Hirano17,Susa19,Inoue20,Chiaki20}.
In addition, turbulence can either promote \citep{Clark11,Riaz18,katharina20}, or suppresses fragmentation due to the amplification of magnetic fields \citep{Sur10,Sur12,Fed11,Turk2012,Machida13,Sharda20}.

Previous studies have shown that the turbulent velocity in minihalos has a similar scaling law to local Galactic molecular clouds \citep{Larson81,Prieto11}.
\cite{Greif2012} demonstrated that the turbulent velocity is amplified during collapse and exceeds the sound speed at the end of the collapse phase.
\cite{Turk2012} also showed that the vorticity of the gas is amplified during the collapse phase.
In addition, they showed that sufficiently high-resolution simulations are required to accurately resolve the morphology of clouds deformed by the turbulence in minihalos. 
However, the driving mechanisms of turbulence have not been well studied.

The results of several cosmological simulations indicate that the accretion flow of gas along the filaments into halos can drive turbulence when dark matter and gas are virialized in minihalos (e.g., \citealt{Wise07,Greif08,Klessen10}).
However, this amplification occurs in low density regions ($n_{\rm H} \sim 10^3~{\rm cm}^{-3}$).
The free-fall time of collapsing dense cores is much shorter, $\lesssim 4.7(n_{\rm H}/10^8 ~{\rm cm^{-3}})^{-1/2}$ kyr, than the dynamical time of the mass accretion onto halos, $\sim 1.5(n_{\rm H}/10^3~{\rm cm^{-3}})^{-1/2}$ Myr,  i.e., the free-fall time scale with averaged density of the gas cloud in the minihalo.
Hence, the accretion-driven turbulence may have negligible effects on the dense cores.
Thus, the turbulence should be amplified by another driving source than the anisotropic gas accretion flow.

\cite{Robertson2012}, \cite{Brinboim18}, and \cite{Mandal20} showed that the turbulent velocity could be amplified by the contraction of isothermal gas clouds. 
This can be explained in the analogy of the {\it inversely} expanding (i.e., contracting) universe. 
In contrast to the cosmological perturbation theory, where the amplitude of velocity fluctuations decays as the universe expands, it should be amplified in collapsing clouds. This mechanism is often quoted as ``adiabatic heating'' of turbulence.

Applying this argument to the central region of clouds hosted by minihalos, we expect that turbulence is amplified by gravitational contraction even with a very weak seed of turbulence at the onset of collapse.  So far, any literature has not shown the detailed comparison of the adiabatic heating theory with numerical simulations of first star formation.  In order to pin down the physical reason of the growth of turbulence in the first star forming core, we need to compare the results of numerical simulations with the theory, including the growth rate, the temporal evolution of the power spectrum, the fraction of the solenoidal modes, etc.


In this paper, we investigate the amplification of turbulence due to gravitational collapse by following gas contraction in minihalos until the end of the collapse phase. The numerical results are analyzed to be compared with the adiabatic heating theory.
This paper is structured as follows. 
First, we analytically estimate a formula for the amplification of turbulence in Section \ref{sec:estimate}. 
In Section \ref{sec:method}, we describe the setup of numerical simulations.
In Section \ref{sec:results}, we present our results and compare them with our analytic model obtained in Section \ref{sec:estimate}, and Section \ref{sec:discussion} is devoted for discussion.
The main points in this paper are summarized in Section \ref{sec:summary}.

\section{Analytic Estimate} \label{sec:estimate}
In this section, we analytically evaluate the amplification of turbulence through gravitational contraction. Gas clouds in minihalos collapse in a self-similar fashion as well as molecular cloud cores in the present-day Universe. Those clouds have a core of Jeans length which contracts approximately uniformly, leaving the outer envelope  with  $ \propto r^{-2.2}$ density distribution (e.g., \citealt{Omukai98,Yoshida06}). Here we focus our discussion on the growth of turbulence in this run-away collapsing core.

We consider the evolution of density/velocity fluctuations in an uniformly changing background.
With the linear approximation, we obtain growth equations for the fluctuations of density $\delta\equiv\delta \rho/{\rho}$ and gravitational potential $\delta \phi$ as 
\begin{eqnarray}
\dot{\delta } + \vb{\nabla} \cdot \vb{u} &=& 0,\label{eq:EoC} \\
\dot{\vb{u}} + 2 H \vb{u} + a^{-2} \vb{\nabla} \delta \phi+\frac{\vb{\nabla}c_{\rm s}^2\delta }{a^2} &=& 0, \label{eq:perturb}\\
a^2\nabla^2 \delta \phi &=& 4\pi G{\rho}\delta.\label{eq:PEq}
\end{eqnarray}
from the equation of motion, the equation of continuity and the Poisson equation.
In these equations, $a$ and $H(=\dot{a}/a)$ denote the scale factor and the ``Hubble parameter'', respectively, that define the change of background density as ${\rho}\propto a^{-3}$.  $c_{\rm s}$ denotes the sound speed,  $\vb{\nabla}$ and $\vb{u}$ are the gradient and peculiar velocity in the comoving frame, respectively. 
Taking the rotation of Equation (\ref{eq:perturb}), we obtain 
\begin{equation}
    \vb{\nabla} \times \dot{\vb{u}} = -2 H \vb{\nabla} \times \vb{u}. \label{eq:rotu}
\end{equation}
Integration of Equation (\ref{eq:rotu}) on both sides yields the scaling relation of the vorticity $\vb{\omega} \equiv \vb{\nabla} \times \vb{u}$ as
\begin{equation}
    \omega \propto a^{-2} \propto \rho^{2/3}. \label{eq:vor}
\end{equation}
This indicates that amplitude of vorticity is proportional to the two-thirds power of the background density.  Hence, the solenoidal mode of turbulence should grow monotonically in the collapsing background.

The relation (Equation \ref{eq:vor}) is valid even in the case that the velocity field is nonlinear.
This can be explained by the fact that the vorticity equation is analogous to the induction equation of magnetic fields,
\[
\frac{\partial \vb{\omega}}{\partial t}=\vb{\nabla} \times (\vb{v} \times \vb{\omega}),
\]
where the baroclinic term and the dissipation term are neglected.
In ideal MHD, it is known that when the magnetic field is frozen with the fluid, the magnetic field increases with contraction by the power of two-thirds of the density, due to the conservation of magnetic flux. 
We can apply the same argument when magnetic field $B$ is replaced with vorticity $\omega$ from the similarity of their growth equations.
In other words, $\omega$ increases due to the conservation of circulation $\Gamma$:
\[
\Gamma \equiv \oint \vb{\omega}\cdot d\vb{S}.
\]
Note that we do not include the viscosity term in the vorticity equation which should play a role when shocks form and the turbulence saturates. Thus, Equation (\ref{eq:vor}) is not valid in the saturated regime. In the numerical simulations of the present paper (Section \ref{sec:method}), the viscosity term is also not included explicitly, since the viscous scale is too small to be resolved. However effective numerical viscosity always comes into play, which realizes the shock formation and the saturation of the turbulence in the simulations.

Similar to the argument on vorticity, we can discuss the divergence of the velocity field, which represents the compression mode of the turbulent flow.
Taking the divergence of Equation (\ref{eq:perturb}) and using Equations (\ref{eq:EoC}) and (\ref{eq:PEq}),  we have

\begin{equation}
\ddot{\delta} + 2H\dot{\delta} -  4\pi G{\rho}-\frac{\nabla^2c_{\rm s}^2\delta }{a^2} = 0,\label{eq:delta_comp}
\end{equation}
where the 3rd and the 4th terms correspond to the self-gravity and the pressure gradient  of density perturbation, respectively. In the present context of our study, we consider the growth of fluctuations in the core of run-away collapsing gas cloud. Hence, we consider the scale below the core radius, which is identical to the Jeans scale. Thus, the 4th term is always larger than the 3rd term which results in the oscillation. The 2nd term behaves as a ``negative drag'' term, which causes the growth of density fluctuations, in contrast to the cosmological context. 

In order to solve this equation, we need to define the collapsing background, $a(t)$. We assume that the core region of the runaway collapsing cloud grows in a self-similar fashion, thereby the density is given as $\rho(t)=5/(12\pi G t^2)$ \citep{Larson69,Suto88}. Note that the origin of the time coordinate is taken when the density approaches infinity (the moment of the protostar formation). The scale factor $a(t)$ is related to the density as ${\rho}(t)=\rho_0a(t)^{-3}$, where $\rho_0$ is the initial density at an initial time $t_0$. After some algebra,  Equation (\ref{eq:delta_comp}) is rewritten as

\begin{equation}
\frac{d^2\delta_k}{d\tau^2} + \frac{4}{3\tau} \frac{d\delta_k}{d\tau} -  \frac{1}{a^3}( 1 -\kappa^2a^{1-3(\gamma_{\rm eff}-1)})\delta_k =0 \label{eq:delta_comp2}
\end{equation}
where  $\delta_k$ is the $k$-mode of density fluctuation,  $\tau\equiv t\sqrt{4\pi G\rho_0}$ is the normalized time coordinate,  $\kappa\equiv c_{s0}k/\sqrt{4\pi G\rho_0}$ is a normalized wave number,  $\gamma_{\rm eff}$  is an effective polytropic index of barotropic equation of state, and $c_{s0}$ denotes the sound velocity at $t=t_0$.  According to the equation of continuity,   $d\delta_k/d\tau$ denotes the divergence of the velocity field,  which gives the compressive mode of the turbulence.
\begin{figure}
 \centering
 \plotone{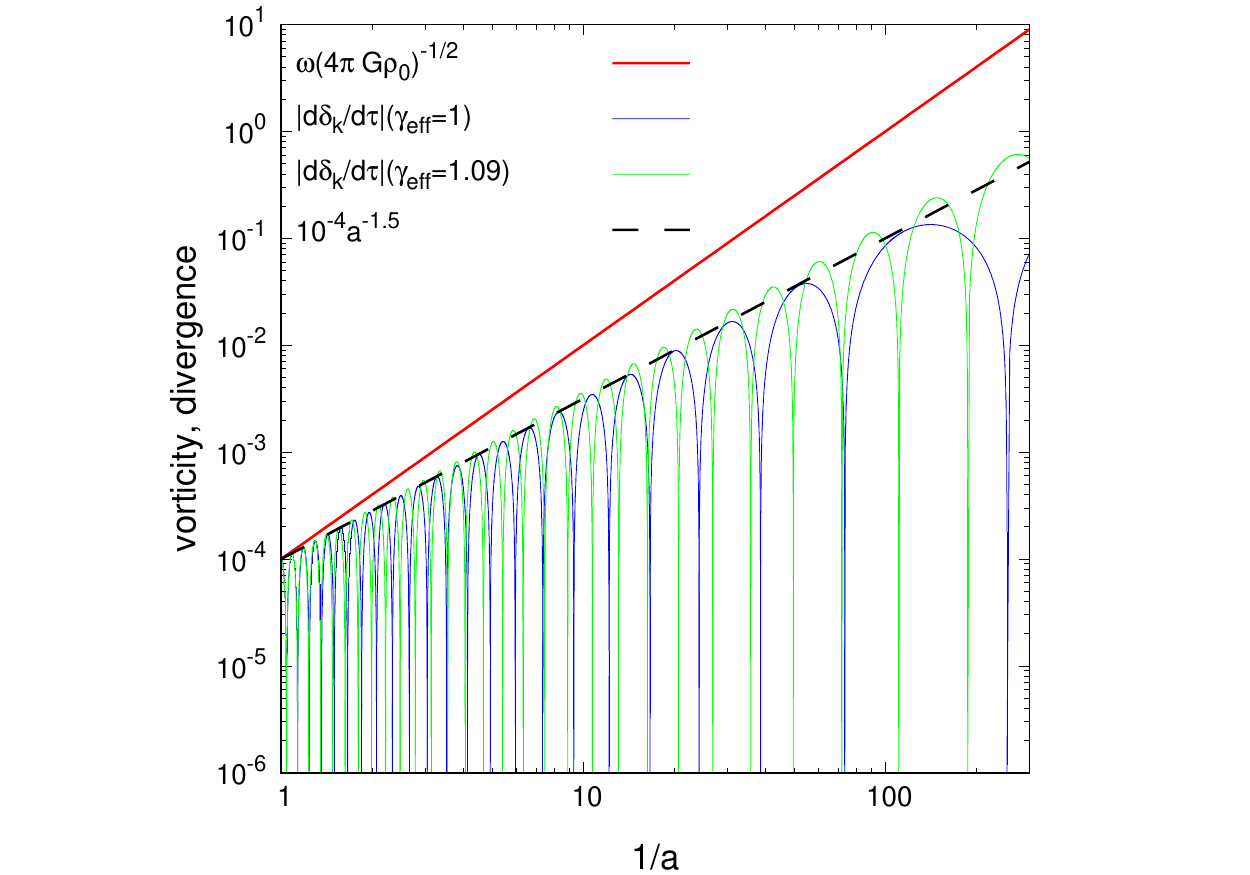}
  \caption{
	Growth of $\omega$ and $|d\delta_k/d\tau|$ is shown for $\kappa =20$.  Red: $\omega$, Blue: $|d\delta_k/d\tau|$ for $\gamma_{\rm eff}=1$, Green: $|d\delta_k/d\tau|$ for $\gamma_{\rm eff}=1.09$.  Dashed line shows $\propto a^{-1.5}$, which traces the amplitude of $|d\delta_k/d\tau|$.
    }
  \label{fig:analytic} 
\end{figure}

Figure \ref{fig:analytic} shows the growth of  $|d\delta_k/d\tau|$ as functions of  $a^{-1}$ by integrating  Equation (\ref{eq:delta_comp2}) for $\gamma_{\rm eff} = 1 $ and $1.09$.  For larger $a^{-1}$ we have higher density, which means that the system evolves from left to right.  For simplicity, the initial condition is set as $\delta_k=0, d\delta_k/d\tau=10^{-4}$.

It is clear that  $|d\delta_k/d\tau|$ oscillates and its amplitude increases as the density gets higher. This overstable behavior is expected, because of the presence of restitutive force term stems from the thermal energy as well as  the external force term by the gravitational contraction of the whole system. Note that the self-gravity term of the perturbation is not important for amplification, because we consider the region inside the collapsing core,  thereby we only have to consider the scale less than the Jeans scale.

The amplitude of the oscillation grows as $\sim a^{-1.5}$ for $\gamma_{\rm eff} =1$ case, and slightly steeper for $\gamma_{\rm eff}=1.09$ (see the dashed line).   This can be shown easily for $\gamma_{\rm eff} =1$ case, since the Equation (\ref{eq:delta_comp2}) can be solved analytically for $\kappa \gg 1$.  The general solution reads
\begin{eqnarray}
\delta_k &=& c_1 j_0(3\kappa(5\tau/3)^{1/3}) +  c_2 n_0(3\kappa(5\tau/3)^{1/3}), \label{eq:bessel1}\\
\frac{d\delta_k}{d\tau} &=& -\kappa(5/3)^{1/3}\tau^{-2/3}\nonumber\\
&\times&\left[ c_1 j_1(3\kappa(5\tau/3)^{1/3}) + c_2n_1(3\kappa(5\tau/3)^{1/3}) \right]. \label{eq:bessel2}
\end{eqnarray}
Here $c_1$ and $c_2$ are constants determined by the initial conditions, $j_0, j_1, n_0, n_1$ are the spherical Bessel functions.  Utilizing the asymptotic form of $j_1(x)\simeq -x^{-1} \cos(x) $ and $n_1(x)\simeq -x^{-1}\sin(x) $ for $x\gg 1 $  and Equation (\ref{eq:bessel2}),  we find that the amplitude of   $d\delta_k/d\tau$ is proportional to $\tau^{-1} \propto a^{-1.5}$.   Remark that this growth rate is less than that of the vorticity $\omega$,  as shown in Fig. \ref{fig:analytic} .

Now we give analytic formulae for the amplification of velocity fluctuations.
%
%
The solenoidal/compressive mode of the turbulent velocity at a given comoving scale $l=2\pi/k$  is 
\begin{eqnarray}
    v(l) =  au \simeq  \begin{cases}
    a \omega l \propto a^{-1}l & {\rm solenoidal}, \label{eq:vel_al}\\
    a\dot{\delta}l \propto a^{-1/2} l & {\rm compressive}.    
   \end{cases}
\end{eqnarray} 
We can regard $v(l)$ as the turbulent velocity. 

As we can see in the above equations  immediately, the compressive mode grows slower than the solenoidal mode. Hence the solenoidal mode dominates the compressive mode, as long as  the initial amplitude of the two modes are at a same order of magnitude\footnote{Note that the expression of Equation (\ref{eq:vel_al}) for the compressive mode is based upon the linear analysis. In case the solenoidal mode becomes much larger than the compressive mode, the nonlinear term becomes important. In such a case, the growth of the compressive mode is boosted by the solenoidal mode.}.  In the line of this, we only consider the solenoidal mode below. 

Since the turbulence consists of the flows with various scales in general, turbulent velocity should be obtained by summing up the contributions from all scales.
In a most simple case, where $l$ is the dominant scale of turbulence, and fixed in the course of the collapse, 
combining Equations (\ref{eq:vor}) , (\ref{eq:vel_al}) and the relation ${\rho} \propto a^{-3}$, we obtain
\begin{equation}
    v \propto {\rho}^{1/3}. \label{eq:vel_cosmo}
\end{equation}
This indicates that the turbulent velocity increases as the gas contracts.

In a realistic case, if the turbulence reaches to the Kolmogorov-like state, $v\propto l^{1/3}$ holds, which means that the turbulent velocity is larger for larger scales. The Jeans length is the core radius, which is the largest scale in the core at a given density. Hence, the Jeans length is the dominant scale. The Jeans length in the comoving frame is given as:
\begin{equation}
    l_{\rm J} \equiv \frac{1}{a}\left( \frac{\pi c_{\rm s}^2}{G {\rho}} \right)^{1/2} \propto {\rho}^{(3\gamma_{\rm eff}-4)/6} \label{eq:jeans}
\end{equation}

Using Equations (\ref{eq:vel_al}) and (\ref{eq:vel_cosmo}) on the scale given by $l = l_{\rm J}$, we may obtain the turbulent velocity at $l_{\rm J}$, but we still need an additional factor.  We need to multiply a factor which comes from the intrinsic shape of the spectrum. 
If we assume the turbulent energy spectrum $E(k)\propto k^{-2\alpha-1}$, the total energy for $k>k_{\rm J}$ is
\begin{equation}
\epsilon(k>k_{\rm J})  =\int_{k_{\rm _J}}^\infty E(k)dk \propto k_{\rm J}^{-2\alpha} \label{eq:ene_alpha}
\end{equation}
where $k_{\rm J}$ equals to $2\pi/l_{\rm J}$, i.e.,  Jeans scale at a given time. 
Using Equations (\ref{eq:vel_al}), (\ref{eq:jeans}), (\ref{eq:ene_alpha}) and $v = \sqrt{2\epsilon}$ we have 
\begin{equation}
    v \propto {\rho}^{1/3} l_{\rm J}^{\alpha} \propto{\rho}^{(3\gamma_{\rm eff}-4)\alpha/6+1/3}. \label{eq:vel_eff}
\end{equation}

For instance, substituting  $\gamma_{\rm eff}=1$ (isothermal) and $\alpha=1/2$ (Larson's law) in the above expression, the exponent becomes $1/4$, and thus the turbulent velocity slowly increases during the collapse.  

Assuming $\alpha$ to be constant, the above relation can be interpreted to the condition that the required initial Mach number ($\mathcal{M}_{\rm 0,cr}$) at $\rho_{\rm 0}$  for the turbulent velocity to be amplified to the level of sound speed at  a given density ($\rho_{\rm sonic}$):
\begin{equation}
\mathcal{M}_{\rm 0,cr} = \left(\frac{\rho_0}{\rho_{\rm sonic}}\right)^{((3\gamma_{\rm eff}-4)\alpha +5 - 3 \gamma_{\rm eff})/6}.
\label{eq:M0cr}
\end{equation}

If we employ $\gamma_{\rm eff}=1.09$ and $\alpha=1/2$,  the critical Mach number is $\sim 0.001$ for $\rho_0=10^{-19} \ {\rm g \ cm^{-3}}$ and $\rho_{\rm sonic}=10^{-6} \ {\rm g \ cm^{-3}}$. 
Considering the initial turbulence of  $\mathcal{M}_{\rm 0,cr}\sim 0.001$ can be easily achieved by accretion flows onto minihalos \citep[e.g.][]{Greif2012},  these analytic estimation suggests that the turbulent velocity in the collapsing core should be account for at least the level of the sound speed when the density approaches to the protostellar density.

\section{Numerical Method} \label{sec:method}

We use the $N$-body/adaptive mesh refinement (AMR) cosmological hydrodynamic simulation code \texttt{Enzo} \citep{Enzo}\footnote{\url{http://enzo-project.org/}} to solve the gas contraction and growth of turbulence.
This code solves the compressive hydrodynamic equations with the Piecewise Parabolic Method (PPM) in an Eulerian frame while using a HLLC Riemann solver.

We use a simplified polytropic model,
\begin{equation}
    P \propto \rho^{\gamma_{\rm eff}},
\end{equation}
where 
$P$ is the gas pressure. 
Initially, we set the uniform gas temperature $T_0 = 200 ~{\rm K}$ and the mean molecular weight $\mu _0 = 1.22$. 
In order to test Equation (\ref{eq:vel_eff}),
we perform simulations with various polytropic indices of
\begin{eqnarray}
\gamma_{\rm eff} = \begin{cases}
    1.2, &  \\
    1.09 & (\text{primordial}), \\
    1.0 & (\text{isothermal}).
  \end{cases}
\end{eqnarray}
The model with  $\gamma_{\rm eff} = 1.09$  mimics the collapse of the primordial gas \citep{Omukai98}.

The initial condition is an isothermal Bonnor-Ebert sphere with an initial peak density $\rho_{\rm peak,0}=4.65\times10^{-20} ~ {\rm g~cm^{-3}}$ which mimics a collapsing primordial gas cloud in the ``loitering'' phase.
To boost the contraction, the density of the Bonnor-Ebert sphere is uniformly enhanced by a factor $f=1.35$ \citep{Matsumoto11}.
The radius $r_{\rm c}$ of the cloud is $1.5 ~ {\rm pc}$, and the size of the computational domain is $L_{\rm box} = 5 ~ {\rm pc}$.
Then, we add the turbulent velocity field. The velocity power spectrum of turbulence is given by the Larson's law $P(k) \propto k^{-4}$ at a wavenumber $k$.
Hence the kinetic energy spectrum $E(k) \propto k^2P(k)$ is proportional to $k^{-2}$ \citep{Dubinski95,Matsumoto15}.
We assume the initial root-mean-square Mach number $\mathcal{M}_0 = 0$, $0.005$, $0.01$, $0.05$, and $0.1$.  
These low Mach numbers are employed because we try to understand the amplification by the gravitational contraction.  
We do not explicitly give the rotational velocity to focus only on the effects of the turbulent velocity field, 
although the turbulence introduces a small amount of angular momentum.

\texttt{Enzo}'s AMR technique allows us to follow the gas contraction over a wide dynamic range.
We start the calculations with a  base grid with $256^3$ computational cells.
We progressively refine cells as the gas cloud collapses so that
the Jeans length is always resolved by at least 128 cells.
Hereafter, the minimum number of cells account for the Jeans length is called as the Jeans Parameter.
The initial conditions are summarized in Table \ref{tab:inits}.

\begin{table}[htb]
 \caption{Initial parameters}
  \centering
   \begin{tabular}{lc}
     \hline \hline
     Parameters & Initial values \\
     \hline
     Central density & $4.65\times10^{-20} ~ {\rm g~cm^{-3}}$  \\ 
     Velocity power spectrum &  $\propto  k^{-4}$   \\
     Mach Number & $0, 0.005, 0.01, 0.05, 0.1$ \\
     Temperature & $200 ~ {\rm K}$   \\
     Mean  molecular weight & 1.22 \\
     Polytropic index & $1.2, 1.09, 1.0$ \\
     Base grid & $256^{3}$ \\
     Jeans Parameter & 128  \\  \hline 
   \end{tabular}
  \label{tab:inits}
\end{table}

In order to investigate the resolution dependence of the amplification of the turbulence, we also run simulations for different numbers of cells on the base grid and Jeans Parameters for $\gamma_{\rm eff}=1.09$.
These models with different base grid sizes and Jeans Parameters are called as RM32--RM512 (Table \ref{tab:reso}).
We conduct RM512 only for $\mathcal{M}_0 = 0.05$  in order to save the computational cost.
\begin{table}[htb]
 \caption{Parameters for resolution study}
  \centering
   \begin{tabular}{ccc}
     \hline \hline
     \multicolumn{2}{l}{Polytropic index} & 1.09   \\ \hline
     Model & Base Grid & Jeans Parameter \\ \hline
     RM32 & $32^3$ &  16 \\
     RM64 & $64^3$ &  32 \\
     RM128 & $128^3$ & 64  \\
     RM256 & $256^3$ & 128 \\
     RM512 & $512^3$ & 256\\ \hline
   \end{tabular} \\
   Note: The other parameters are same as Table \ref{tab:inits}.\\
   RM512 is only for $\mathcal{M}_0=0.05$.
  \label{tab:reso}
\end{table}



We terminate the simulations when the peak density exceeds $10^{-4} ~{\rm g~cm^{-3}}$ for all models. 
We use the \textsc{yt} toolkit \citep{yt}\footnote{\url{https://yt-project.org/}} to analyze the data for all simulations.
\section{Results} \label{sec:results}
\subsection{Overall density structure}\label{ssec:density}
\begin{figure*}[htbp]
 \centering
 \plotone{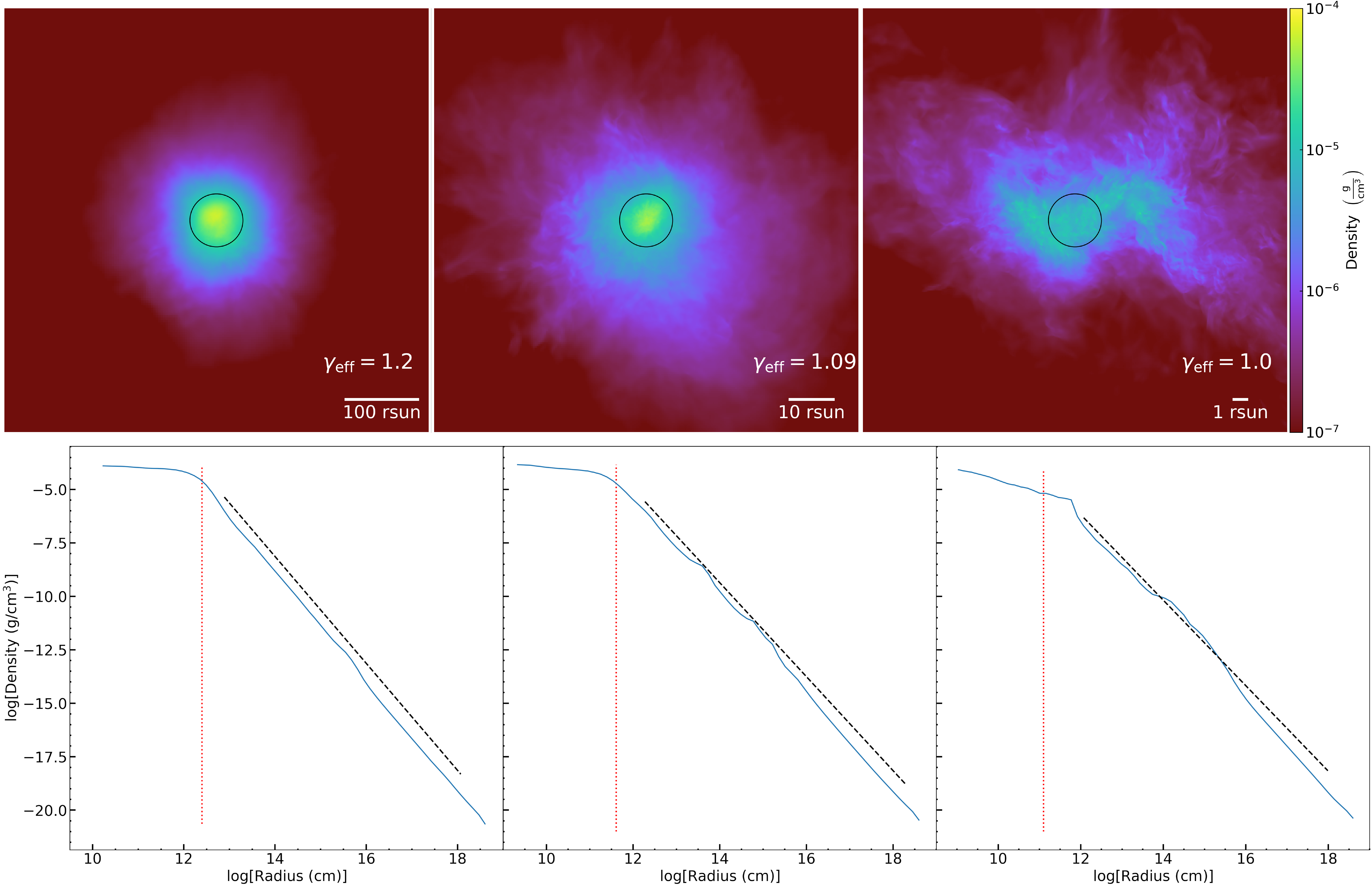}
  \caption{
  Density-weighted density projection plots (upper panel) and radially averaged profiles of the density (lower panel) for the model with $\mathcal{M}_0=0.1$ when the peak density reaches $10^{-4} ~ {\rm g~cm^{-3}}$. In the upper panel, side length of each plot is set to $8 L_{\rm J}$ and the black circles have a radius of $L_{\rm J}/2$.
  The red dotted lines in the lower panel indicate $L_{\rm J}/2$.
  The black dashed lines in the lower panel denote $\propto r^{-2/(2-\gamma_{\rm eff})}$, where $r$ is the radius from the center of the gas clouds.
  }
  \label{fig:projection} 
\end{figure*}

Figure \ref{fig:projection} shows the density-weighted density projection plots (upper panel) and radially averaged profiles of the density (lower panel) for the model with $\mathcal{M}_0=0.1$ when peak density reaches $10^{-4} ~ {\rm g~cm^{-3}}.$
The self-similar evolution of the gas cloud forms the core-envelope structure in the density distribution. The density-radius relation in the envelope is consistent with the analytic expression $\rho \propto r^{-2/(2-\gamma_{\rm eff})}$ \citep{Suto88}, as shown in the lower panel.

We can see the disturbed density fields and deformation of the gas clouds for $\gamma_{\rm eff}=1.09$ and $1.0$, while the density distribution is less fluctuated and spherical for $\gamma_{\rm eff}=1.2$. This is due to the gravitational instability of the system for cloud deformation.
These results are consistent with the semi-analytic studies, which show bar-mode perturbations {grow on spherically collapsing clouds} with $\gamma_{\rm eff} < 1.097$ \citep{Hanawa00,Lai2000}.

\begin{figure}[htbp]
 \centering
 \plotone{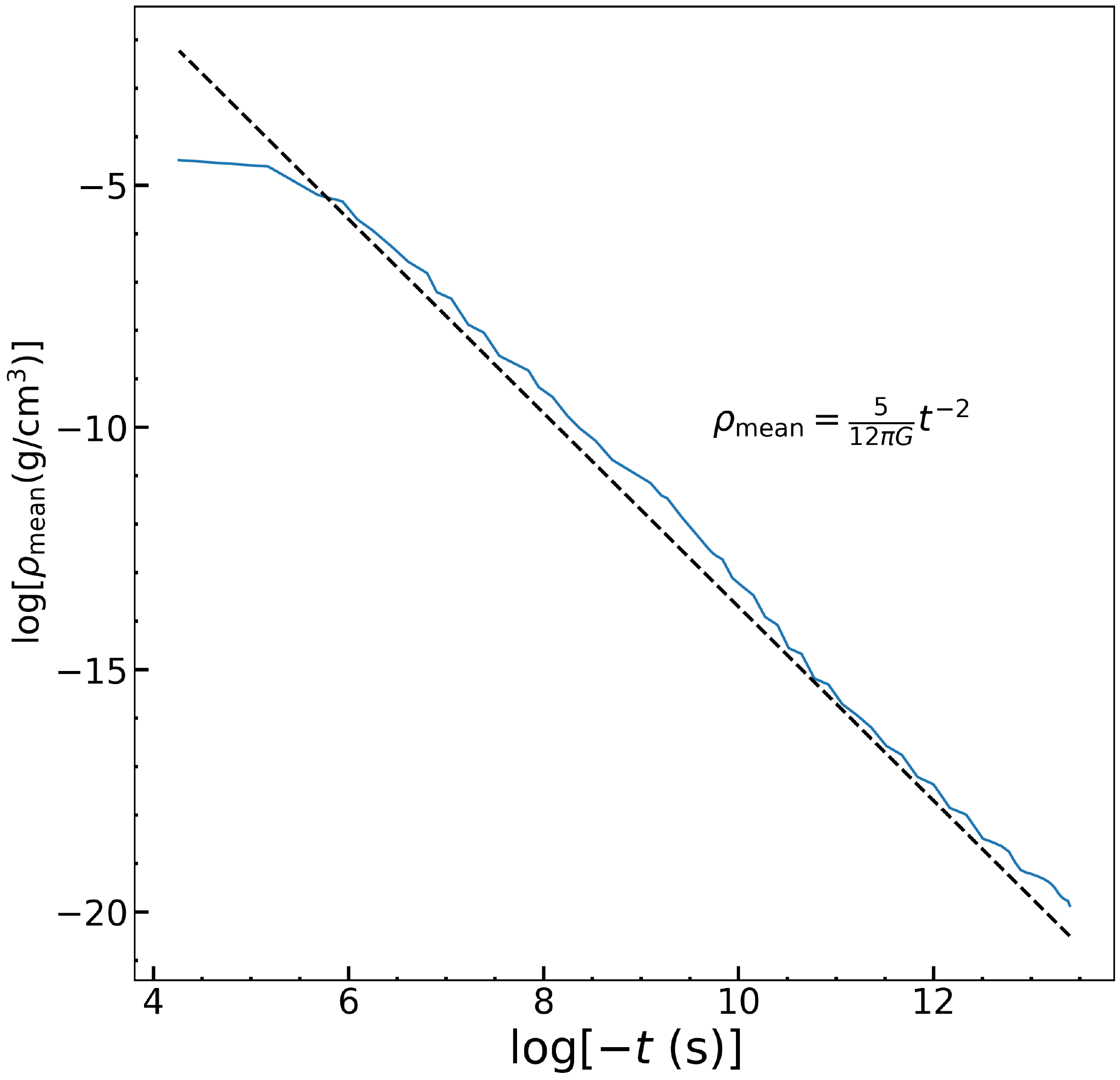}
  \caption{Evolution of the mean density of the core as a function of time for $\mathcal{M}_0=0.1$ and $\gamma_{\rm eff}=1.09$.
  A black dashed line corresponds to $\rho_{\rm mean} = \frac{5}{12 \pi G }t^{-2}$.}
    \label{fig:time_dens} 
\end{figure}

Figure \ref{fig:time_dens} shows the evolution of the mean density of the core as a function of time $-t$ for $\mathcal{M}_0=0.1$ and $\gamma_{\rm eff}=1.09$.
Here the origin of the time coordinate is at the time of final snapshot, which approximately represents the time when the core density diverges in the similarity solution.
The mean density $\rho_{\rm mean}$ is calculated in a spherical region  with a radius $L_{\rm J}/2$ (hereafter called ``Jeans volume'') of every snapshot. Here $L_{\rm J}$ is the Jeans  length in the proper coordinate.

We can clearly see the mean density evolves following the relation $\rho_{\rm mean} = \frac{5}{12 \pi G }t^{-2}$.
This again agrees with the Larson-Penston type similarity solution for a gravitationally contracting polytropic sphere \citep{Suto88}.
These results support our assumption for the analytic estimate in Section \ref{sec:estimate}.

\subsection{The growth of turbulence}\label{ssec:highest}
We calculate the average turbulent velocity in the Jeans volume.
The turbulent velocity $v_{\rm turb}$ is defined as
\begin{equation}
    v_{\rm turb}^2\equiv \sum_{\leq L_{\rm J}/2 } \frac{V_i}{V_{\rm J}}(\vb{v}_i-\vb{v}_{{\rm rad},i})^2, \label{eq:vturb}
\end{equation}
where $V_i$ is the volume of the $i$'th cell, and $V_{\rm J}$ is the Jeans volume. 
It is difficult to estimate the {\it unperturbed} radial velocity ${\vb v}_{{\rm rad}, i}$ at the position of each cell $i$ because the velocity fluctuation is contaminated with the background radial velocity in the simulation data.
We guess the background radial velocity to estimate the turbulent velocity field by means of the following procedure:
First, we calculate the radial velocity profile with $N_{\rm rad}$ radial bins.
$N_{\rm rad}$ should be smaller than the Jeans parameter so that we can remove the velocity fluctuations at the scale of the cell size.
Here, we set $N_{\rm rad} = 16$.
Second,  we linearly interpolate this profile to estimate the smoothed radial velocity $\vb{v}_{{\rm rad}, i}$ at the position of each cell.
Third, we subtract $\vb{v}_{{\rm rad}, i}$ from the total velocity $\vb{v}_{\rm i}$ of the cell to extract the component of the fluctuation.
Finally, we calculate the cell volume-weighted average of the turbulent velocity among cells in the Jeans volume according to Equation (\ref{eq:vturb}).
We use the `volume-weighted' average instead of the `mass-weighted' average here. 
We do not find any significant difference between the results from the two methods because we are now considering only the central part of the gas cloud, where the density inhomogeneity is not so large {even in the isothermal model($\sqrt{\langle \delta^2 \rangle }\lesssim 0.9$).}  We will discuss the density fluctuation in section \ref{sec:density_fl} and in Figure \ref{fig:delta-t}.

\begin{figure}[htbp]
 \centering
  \includegraphics[scale=0.4]{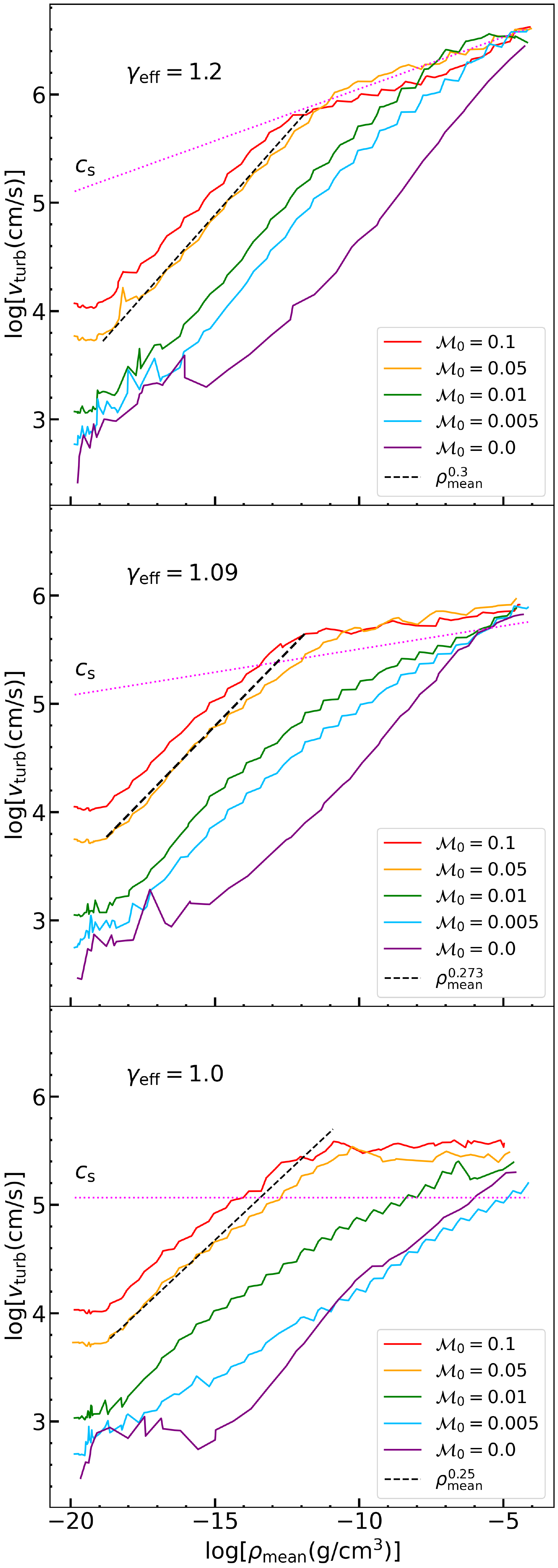}
  \caption{
    Evolution of the turbulent velocity $v_{\rm turb}$ as a function of mean density $\rho _{\rm mean}$ in a spherical region centered on the density mean with a radius of half of the Jeans length (Jeans volume) for $\gamma_{\rm eff}=1.2$ (top panel), $1.09$ (middle panel), and $1.0$ (bottom panel). The solid curves with different colors denote the numerical results for different initial turbulent Mach numbers $\mathcal{M}_0$. The magenta dotted line denotes the average sound speed in the Jeans volume. The black dashed lines depict our analytic estimates (Equation \ref{eq:vel_eff}).}
  \label{fig:vel_p} 
\end{figure}

\begin{figure}[htbp]
 \centering
 \plotone{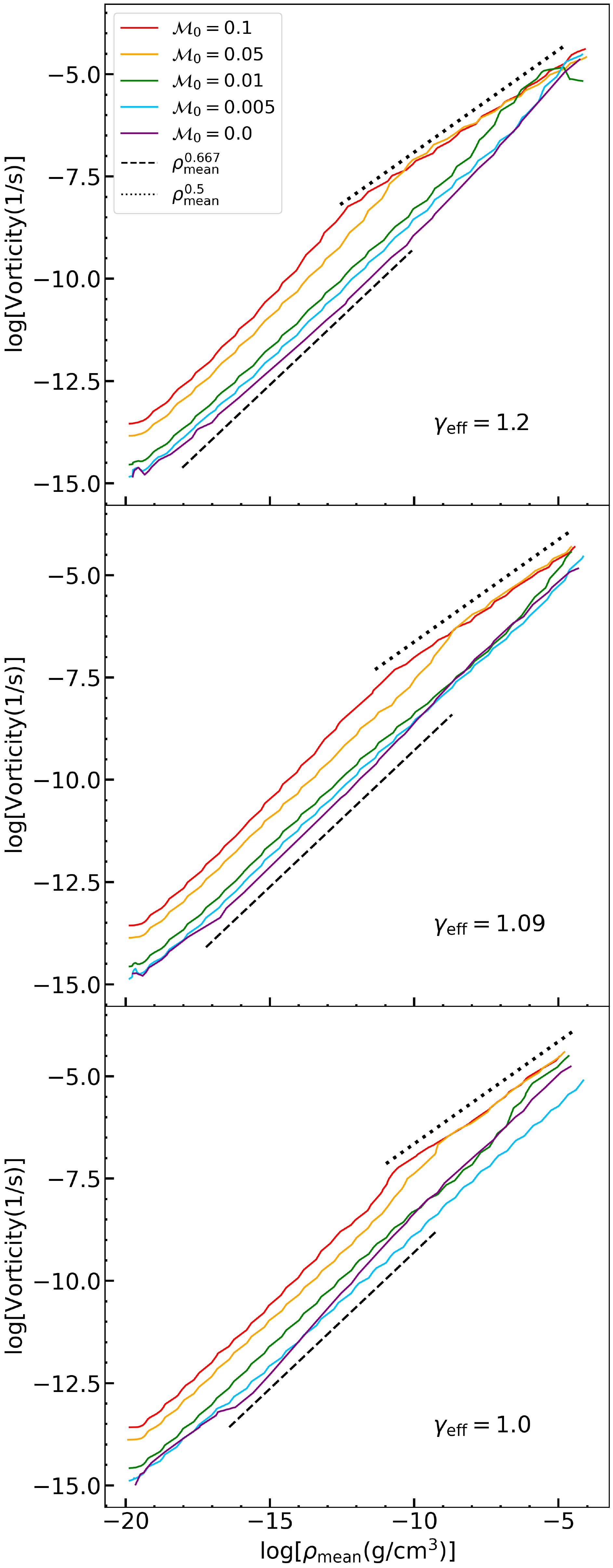}
  \caption{
    Evolution of the vorticity as a function of the mean density for
    $\gamma_{\rm eff}=1.2$ (top panel),
    $1.09$ (middle panel), and
    $1.0$ (bottom panel).
    The solid curves denote the numerical results of each $\mathcal{M}_0$.
    The black dashed line is our analytic estimate (Equation \ref{eq:vor}).
    {The black dotted line denotes $\propto \rho_{\rm mean}^{0.5}$.} 
    }
    \label{fig:vor_p} 
\end{figure}

Since the mean density $\rho _{\rm mean}$  of the Jeans volume almost monotonically increases, we can show the temporal evolution of the turbulent velocity $v_{\rm turb}$ 
as a function of $\rho _{\rm mean}$ of every snapshot in Figure \ref{fig:vel_p}.
In all models, the turbulent velocity increases as the density increases through the contraction.
Similarly, Figure \ref{fig:vor_p} shows the evolution of the vorticity as a function of the mean density for various initial Mach numbers.
In all models, the vorticity increases monotonically with the increasing density. 

In these figures, for $\mathcal{M}_0 \geq 0.05$, we can see that Equation (\ref{eq:vel_eff})  and  Equation (\ref{eq:vor}) are in good agreement with the numerical results for all $\gamma_{\rm eff}$ before they reach the sound velocity. We also find that all models with $\mathcal{M}_0 \geq 0.05$ reach to the transonic/supersonic points before it reaches the protostellar density ($\rho_{\rm peak} \gtrsim 10^{-4} ~ {\rm g~cm^{-3}}$ ).  Hence, the run-away collapsing  core will easily achieve the turbulent velocity of $\sim c_{\rm s}$, even with a weak initial turbulence. In fact, $\mathcal{M}_0$ of these models is larger than  the threshold Mach number shown in  Equation (\ref{eq:M0cr}) for $\rho_{\rm sonic}=10^{-6} ~{\rm g~cm^{-3}}$ and $\rho_0= 4.65\times 10^{-20} ~ {\rm g~cm^{-3}}$.
Considering the good match with the theory in Section \ref{sec:estimate} with the numerical results, this is a good evidence that the turbulent velocity is amplified  via adiabatic heating associated  with the  gravitational contraction.
%

It is worthy to note that the turbulent velocity is saturated at $\sim c_{\rm s}$ for $\gamma_{\rm eff}=1.2$ models, whereas it reaches well above the sound velocity in other two models (Figure \ref{fig:vel_p}).  
 These results naturally explain the presence of supersonic turbulence in the first star forming core in cosmological simulations \citep[e.g.][]{Greif2012}.
In Figure \ref{fig:vor_p}, we also find the sign of saturation for all $\gamma_{\rm eff}$ for $\mathcal{M}_0 \geq 0.05$. 
The vorticity after the saturation roughly obeys $\omega \propto \rho_{\rm mean}^{1/2}$, which is predicted by \cite{Robertson2012}.  They also have shown that uniformly contracting isothermal gas ($\gamma_{\rm eff}  = 1 $) with initially supersonic turbulence ($ \mathcal{M} _0=6$) can reach as high as $ \mathcal{M} =11.2$. This Mach number is much higher than that is found in our simulations. We guess the main reason of this difference is that they did not solve the gravitational collapse, but we do. They mimic the collapse by increasing the average density at a given rate by hand, like ``collapsing universe''. {In our calculation, the cloud core size shrinks as the collapse proceeds, so we guess the decrease of the forcing scale (the core size, $L_{\rm J}$ in our case) causes the higher dissipation rate of turbulence (see eq. 7 in \cite{MacLow1999}). We speculate this rapid dissipation avoids highly supersonic turbulent flows}.


For $\mathcal{M}_0 \leq 0.01$,
the turbulent velocity oscillates around $\sim 0.01c_s$ at the early stage of the collapse,  ($\rho_{\rm mean} < 10^{-15}~{\rm g~cm^{-3}}$) as shown in Figure \ref{fig:vel_p}.  
As will be discussed in Section \ref{ssec:resolution},  this behavior implies that the turbulent velocity at this stage is contaminated by the numerical error introduced by the Cartesian grid. The initial seed velocity field is so weak that it is hidden among the added errors due to grid discretization.   
After this early stage, as the eddies  being resolved, the turbulent velocity turns to increase even slightly faster than the cases for $\mathcal{M}_0 \geq 0.05$, {roughly proportional to $\rho_{\rm mean}^{1/3}$.} This may be due the flattening of the energy spectrum around the Jeans scale.
This point will be discussed in Section \ref{ssec:mode}.
Finally, the turbulent velocities in all models reaches to the order of sound velocity.  However, as mentioned above, the turbulent velocity in the models of $\mathcal{M}_0 \leq 0.01$ are floored by numerical errors. Hence these results seems to be a numerical artifact. 
On the other hand, the initial Mach number of these models except $\mathcal{M}_0=0.0$ exceeds the critical number described as Equation (\ref{eq:M0cr}). Thus, it is expected that these models will achieve the transonic/supersonic turbulence even if enough resolution is provided.

\begin{figure}[htbp]
 \centering
 \plotone{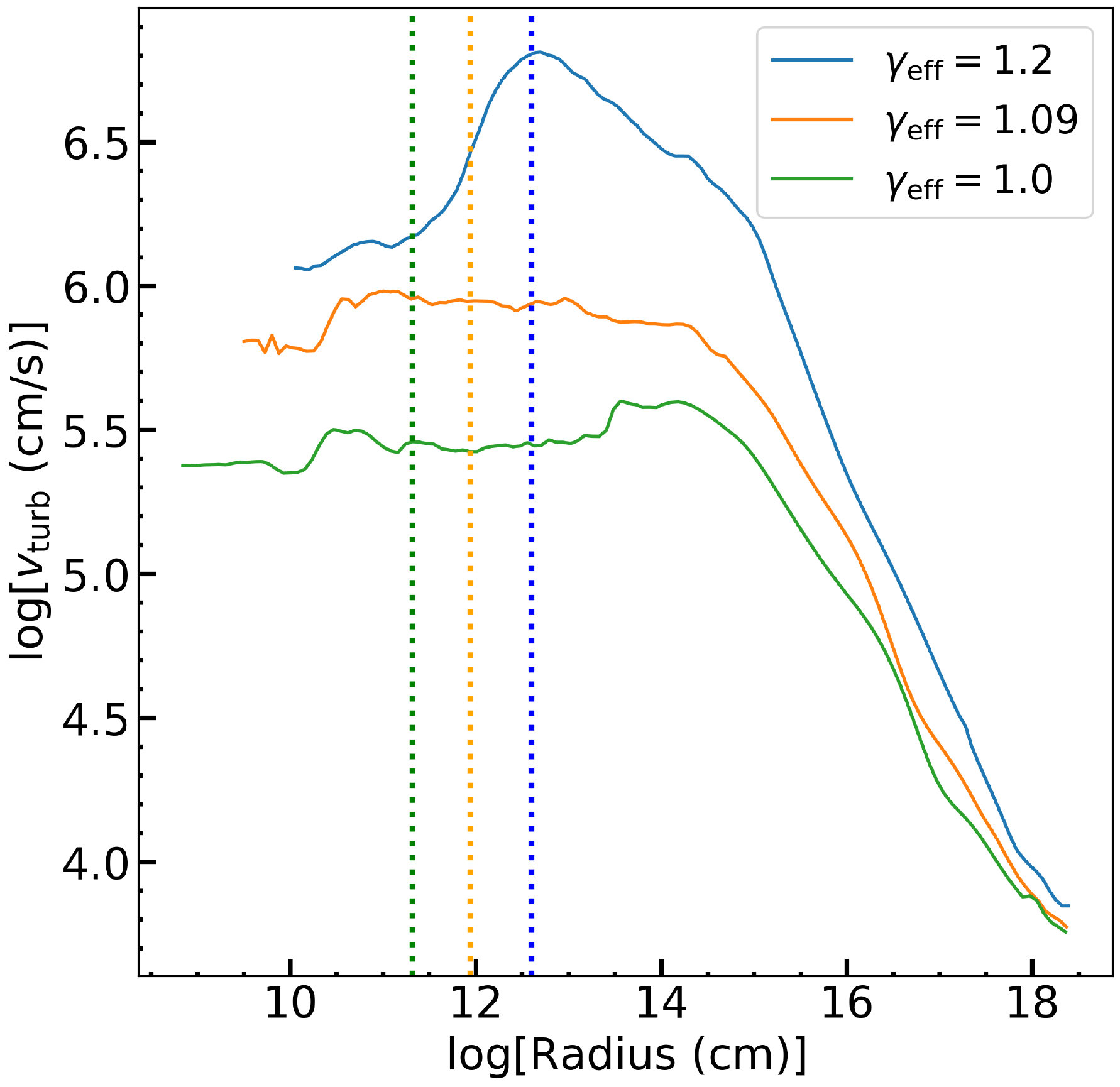}
  \caption{
    Radial profiles of turbulent velocity averaged over the shell of radial bins with $\mathcal{M}_0=0.05$ for each $\gamma_{\rm eff}$ when the peak density reaches $10^{-4} ~ {\rm g ~ cm^{-3}}$.
    {The vertical lines correspond to the half of Jeans length for each $\gamma_{\rm eff}$.}}
    \label{fig:radturb} 
\end{figure}

We also plot the radial profiles of turbulent velocity averaged over the shell of radial bins as a function of radius from the center of mass of the cloud in Figure \ref{fig:radturb}. This plot is for $\mathcal{M}_0 = 0.05$. Since the turbulent velocity increases with contraction, it exhibits a core-envelope structure as the density distribution shows. 
In the core region, the turbulence is saturated, and its amplitude seems to be determined by the saturation level (Figure~\ref{fig:vel_p}).
The saturation occurs at densities $\sim 10^{-8} \ {\rm g/cm^{3}}$, much smaller than the peak density. Therefore, the core radius in the turbulent velocity profile is larger than the core in the density profile.
We also find that the radial profile tends to drop towards the cloud center. We do not understand the actual reason of this drop, but the profile of $\gamma_{\rm eff}=1.09$ is more or less consistent with Fig.1 in \cite{Fed11} who employ $\gamma_{\rm eff}=1.1$.

\begin{figure}[htbp]
 \centering
  \includegraphics[scale=0.30]{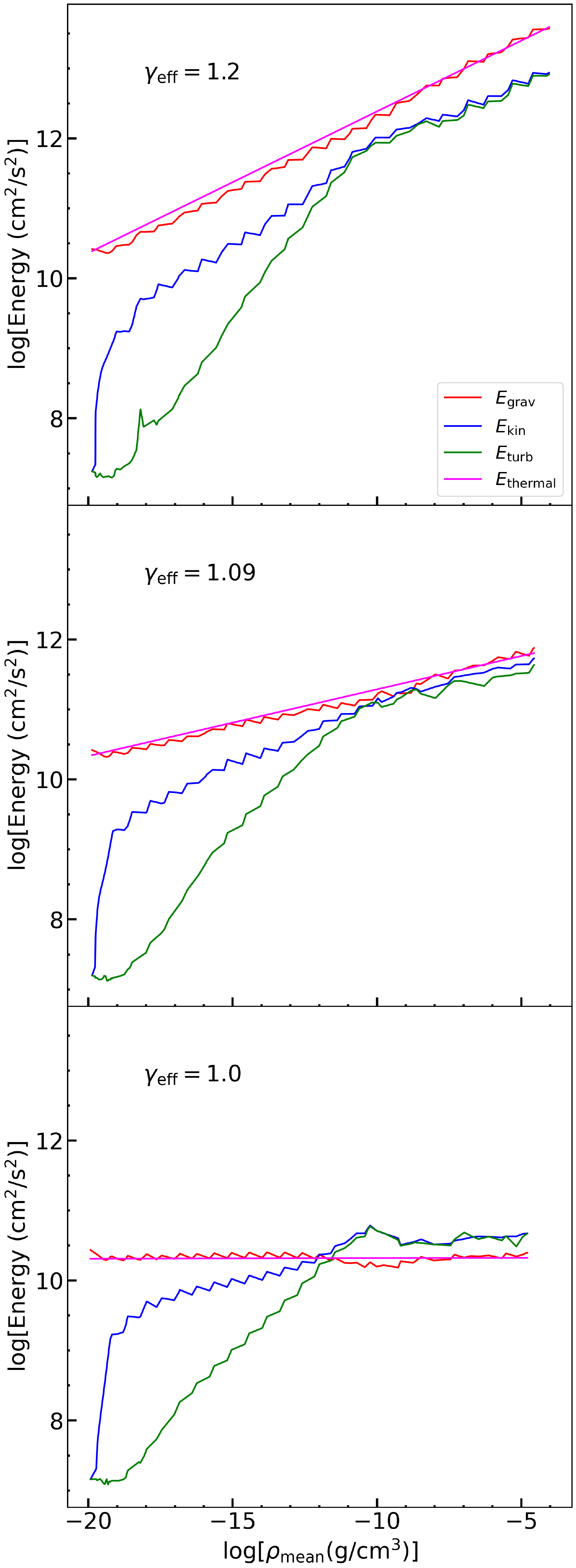}
  \caption{
    Evolution of the specific energy for $\mathcal{M}_0=0.05$ as a function of the mean density for
    $\gamma_{\rm eff}=1.2$ (top panel),
    $1.09$ (middle panel), and
    $1.0$ (bottom panel).
    Red, blue, green, and magenta lines denote the gravitational, kinetic, turbulent, and thermal specific energies, respectively. 
    }
    \label{fig:energy} 
\end{figure}

Figure \ref{fig:energy} shows the evolution of gravitational, kinetic, turbulent, and thermal specific energies as a function of mean density.

Here, `turbulent energy' means the kinetic energy minus the energy of bulk radial motion.
{We define each specific energy to compare in the same units as follows: 
\begin{eqnarray}
  E_{\rm grav} \equiv \left|- \frac{3}{5}\frac{GM_{\rm J}}{L_{\rm J}/2}\right|, ~~~ E_{\rm kin} \equiv \frac{1}{2}v^2, \nonumber \\
  E_{\rm turb} \equiv \frac{1}{2}v_{\rm turb}^2, ~~~ E_{\rm th} \equiv \frac{1}{\gamma_{\rm ad}-1}\frac{k_{\rm B}T}{\mu m_{\rm H}}. \nonumber
\end{eqnarray}
Here, $M_{\rm J}$, $\gamma_{\rm ad}$, $k_{\rm B}$, and $m_{\rm H}$ are mass within the Jeans length, the specific heat ratio of the gas, Boltzmann constant, and hydrogen mass, respectively.}

To calculate the thermal energy we assume {$\gamma_{\rm ad}$} to 5/3 because we do not solve chemical reactions in this work.
{The factor 3/5 in $E_{\rm grav}$ comes from the approximation that the core is uniform. $E_{\rm kin}$, $E_{\rm turb}$ and $E_{\rm th}$ are evaluated as the cell volume weighted averages, again with the uniform core approximation.} 
For $\gamma_{\rm eff}$=1.0, the kinetic and turbulent energies increase until they are comparable to the gravitational energy, while in the other models they increase but do not reach that level.
The detailed physical mechanism of saturation and the dependence of the saturation level on $\gamma_{\rm eff}$ are interesting, but they are out of scope of this work.
We will address these issues in our future study.

\subsection{Solenoidal/Compressive modes}\label{ssec:mode}
%
%
The turbulent field is composed of solenoidal and compressive modes. The divergence free velocity fields corresponds to the solenoidal mode, and the rotation free fields are compressive. As we discussed in Section \ref{sec:estimate}, the turbulent velocity which is amplified by the contraction is accompanied by vorticity and divergence.  Thus, it is interesting that which mode is dominant for the amplification of the turbulence. We study this point in this section. We are also interested in the energy spectrum, because Equation (\ref{eq:vel_eff}) critically depends on the assumption that spectrum is described by a constant single power law index $\alpha$, which equals to 1/3 for Kolmogorov turbulence and equals to 1/2 for Larson's law. In this section, we focus on the results in the model of $\gamma_{\rm eff}=1.09$.

We compute the velocity in Fourier space (hereafter $k$-space) on a uniform grid compatible with the highest refinement level, using a fast Fourier transform with a window of the spherical Bessel function in the cube of $L_{\rm J}$ on one side.
The reason why we use the window function is to suppress the ``side-lobe'' caused by the non-periodic boundary condition of the cube.
Applying the Helmholtz decomposition, we can decompose the obtained velocity field into solenoidal modes $\hat{\vb{v}}_{\rm sol}(k)$ and the compressive modes $\hat{\vb{v}}_{\rm comp}(k)$ in $k$-space as
\begin{eqnarray}
    \hat{\vb{v}}_{\rm comp}(\vb{k}) &=& (\vb{k}\cdot\hat{\vb{v}}(\vb{k}) )\vb{k}/k^2, \nonumber \\
    \hat{\vb{v}}_{\rm sol}(\vb{k})   &=&
    (\vb{k}\times\hat{\vb{v}}(\vb{k}))\times\vb{k}/k^2,
\end{eqnarray}
where $\vb{\hat{v}(\vb{k}})$ is the velocity in $k$-space and $\vb{k}$ is the wave vector.
The minimum wavenumber corresponds to the spatial scale of half of the Jeans length.
Then, we calculate the kinetic energy of each mode as
\begin{eqnarray}
    E_{\rm comp}&=&\int_{\rm k_{\rm J}}^\infty\hat{E}_{\rm comp}(k)dk \equiv  \frac{1}{2k_{\rm J}^3} \int_{k_{\rm J}}^\infty \left|\hat{\vb{v}}_{\rm comp}(\vb{k})  \right|^2 4 \pi k^2 dk , \nonumber \\
    E_{\rm sol}&=&\int_{\rm k_{\rm J}}^\infty\hat{E}_{\rm sol}(k) dk   \equiv  \frac{1}{2k_{\rm J}^3} \int_{\rm k_{\rm J}}^\infty\left|\hat{\vb{v}}_{\rm sol}(\vb{k}) \right|^2 4 \pi k^2dk .
\end{eqnarray}

\begin{figure}
  \centering
  \plotone{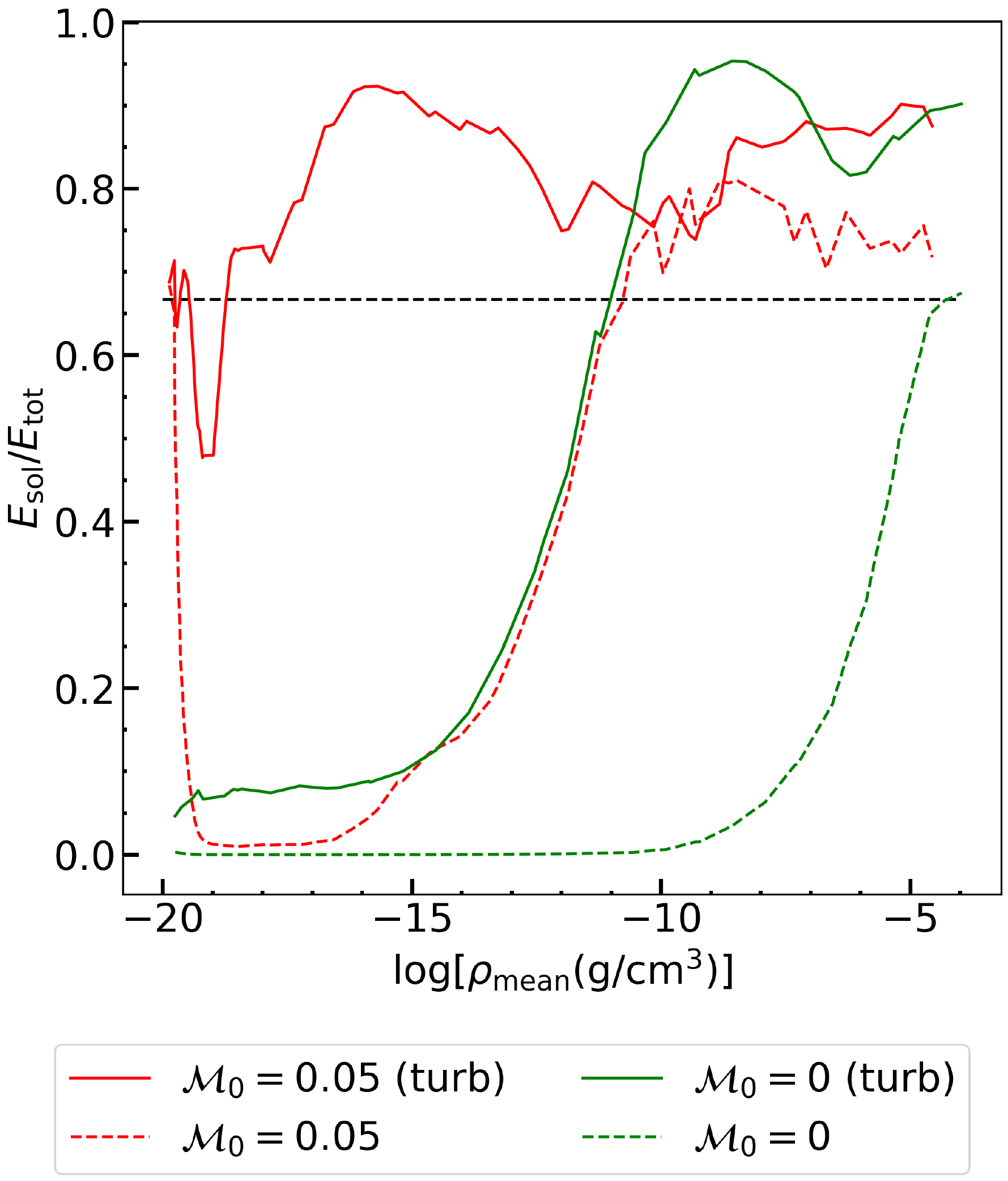}
  \caption{
    Evolution of the solenoidal ratio as a function of the mean density.
    The solid lines denote the solenoidal ratio of the turbulent velocity for each model of $\mathcal{M}_0$.
    The dashed lines are the same as the solid lines, but the cases of the total velocity.
    The black dotted line corresponds to the natural mixture ratio.
    $\gamma_{\rm eff} = 1.09$ in every model.
    }
  \label{fig:ratio} 
\end{figure}

First, we show the growth of the energy of solenoidal mode in Figure \ref{fig:ratio}. 
We define the solenoidal ratio as $E_{\rm sol} / E_{\rm tot}$ where $E_{\rm tot}$ is the sum of solenoidal and compressive modes.
The black dotted line in Figure \ref{fig:ratio} shows the solenoidal ratio for the natural mixture.
As discussed in \cite{Fed11}, this natural ratio $\sim 2/3$ can be approximated to be the number ratio of the modes of longitudinal waves which induce the rotational motion of the gas in a three-dimensional system. 
The dashed curves of Figure \ref{fig:ratio} show the total velocity including the bulk motion, whereas the solid curves denotes the turbulent velocity subtracted the radial infall $\vb{v}_{{\rm rad},i}$  defined in Section \ref{sec:method}.
For the model with $\mathcal{M}_0=0.05$,  the solenoidal ratio is initially close to 0.67, because we initially set the velocity field with the natural mixture.
In this case, the solenoidal ratio of the turbulent component (solid curves) increases up to $\sim 0.9$ after the collapse begins, and maintains its high value until the end of collapse.
This is because the growth rate of the solenoidal mode is larger than that of the compressive mode.
\begin{figure*}
  \centering
  \includegraphics[scale=0.4]{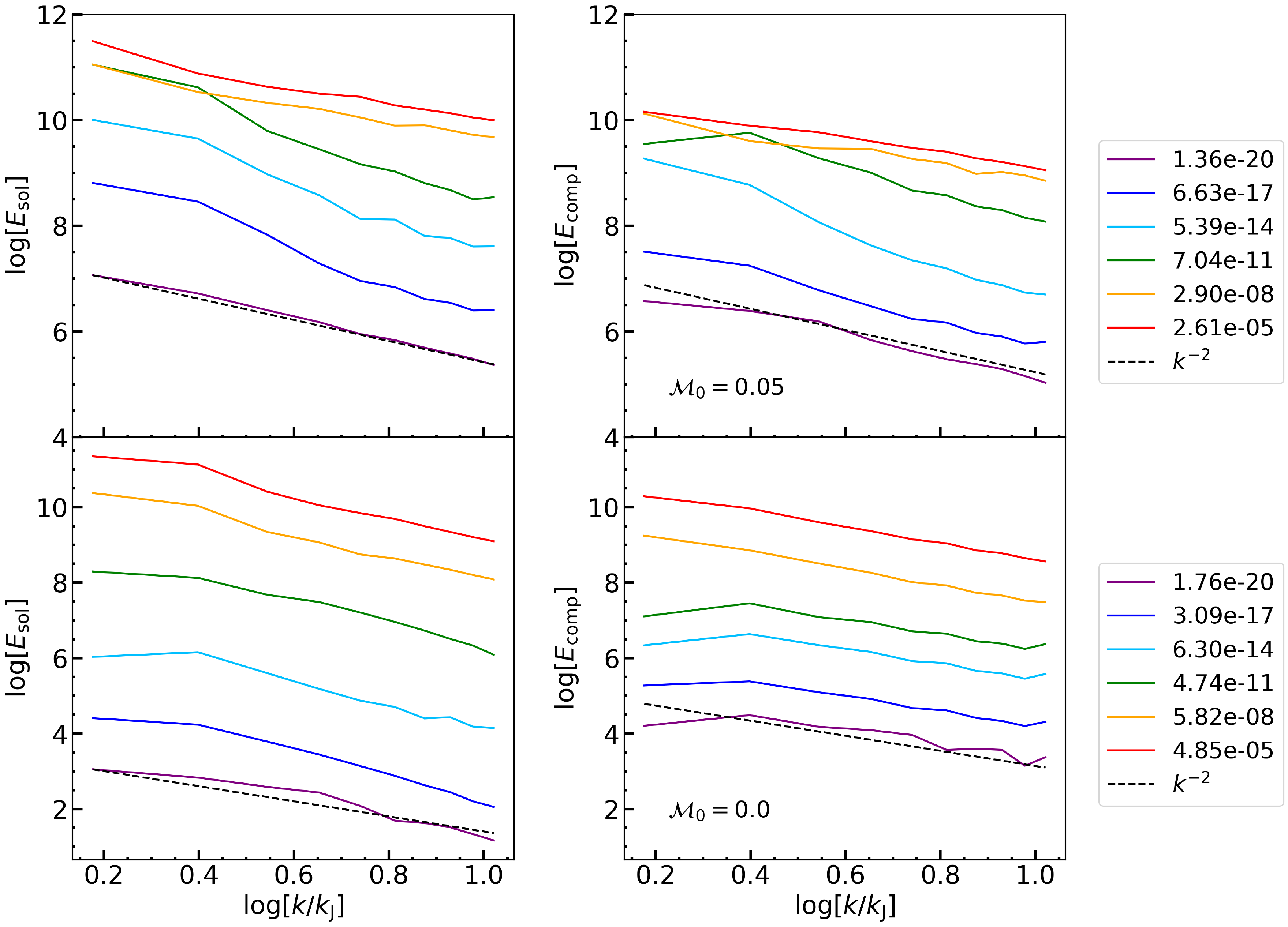}
  \caption{
    Evolution of the turbulent kinetic energy spectra of solenoidal modes (left column) and compressive modes (right column) for $\mathcal{M}_0=0$ (upper row) and $0.05$ (lower row), respectively.
    The solid curves show the numerical results.
    The dashed curves show reference power-law spectra $\propto k^{-2}$, indicating the scaling relations of the Larson's law turbulence.
    {The different colors indicate different values of $\rho_{\rm mean}$.}
    }
  \label{fig:spectrum} 
\end{figure*}
In short, for model $\mathcal{M}_0=0.05$, the turbulence is dominated by the solenoidal mode, while it is growing.

In contrast to this case, the solenoidal ratio is relatively low in the low-density region for the model with $\mathcal{M}_0=0$. This behavior suggests that the numerical error introduced in the low density region is basically the compressive mode.
This is reasonable, because the error accrues concomitant with the radial converging flow. As the collapse proceeds, the solenoidal ratio reaches $\sim 0.9$ in the high-density region as well as for $\mathcal{M}_0=0.05$ (green and red solid curves). This means that the launch of the growth of solenoidal mode is delayed for $\mathcal{M}_0 = 0$, but eventually overwhelms the compressive mode.

For the kinetic energy including the radial inflow velocities (dashed curves), the solenoidal ratio decreases rapidly after the onset of collapse, and almost close to zero in the both cases. As the gas collapses, the solenoidal ratio gradually increases due to the amplification of the solenoidal modes, and eventually converges to the natural mixture, which is consistent with the results of \cite{Fed11}.

Next we show six snapshots of the  energy spectrum in Figure \ref{fig:spectrum}.
Left panels show the solenoidal mode, and the right panels show the compressive mode. Upper and lower row denote the $\mathcal{M}_0=0.05$ and $\mathcal{M}_0=0$ models, respectively. {The different colors indicate different values of $\rho_{\rm mean}$.} The horizontal axes show the wave number normalized by the Jeans scale at {the corresponding mean density}.

For $\mathcal{M}_0 = 0.05$, the solenoidal mode roughly keeps a power law energy distribution.
Initially, it is set as $\propto k^{-2}$ and the slope at small $k$ end ($< 5 k_{\rm J}$) keeps $\propto k^{-2}$.

It is also important that $E_{\rm sol}$ basically a few--10 times larger than $E_{\rm comp}$ at each snapshot.  

On the other hand, for $\mathcal{M}_0 = 0 $ model, compressive mode dominates initially, although  no turbulence is added by hand.\footnote{ Note that this earliest snapshot corresponds to the time slightly after the start of the simulation, not at exactly the {beginning of} the calculation. }
This is obviously the error introduced by discretization, that should accompany the radial inflow. 
As the collapse proceeds, the solenoidal mode takes over the compressive mode at $\rho_{\rm mean} \gtrsim 10^{-11} ~{\rm g~cm^{-3}}$. The initial growth of solenoidal mode (bottom 3--4 curves)  is remarkable, that seems to be boosted by the large compressive mode. The boost by the compressive mode ends earlier for smaller $k$ eddies and the growth becomes slower. This is because $E_{\rm sol}$ for small  $k$ mode is larger than that of large $k$ modes, to overtake the compressive mode earlier. 
As a result, $E_{\rm sol}$ for larger $k$ mode are able to catch up that of smaller $k$ modes. This effect causes the flattening of the spectrum of $E_{\rm sol}$, which we observe in the spectrum from the second (blue) to fourth (green) snapshots. In particular, the low $k$ end of the fourth snapshot is nearly flat. This means $\alpha \simeq 0$ in Equation (\ref{eq:vel_eff}) to have $v\propto \rho^{1/3}$ for $\gamma_{\rm eff}=1.09$. This trend is already observed in the middle panel of Figure \ref{fig:vel_p}. However, as we already have mentioned, this growth relies on the presence of the numerical error in the compressive mode. Hence this should be regarded as a numerical artifact. 

\begin{figure}[htbp]
  \centering
  \plotone{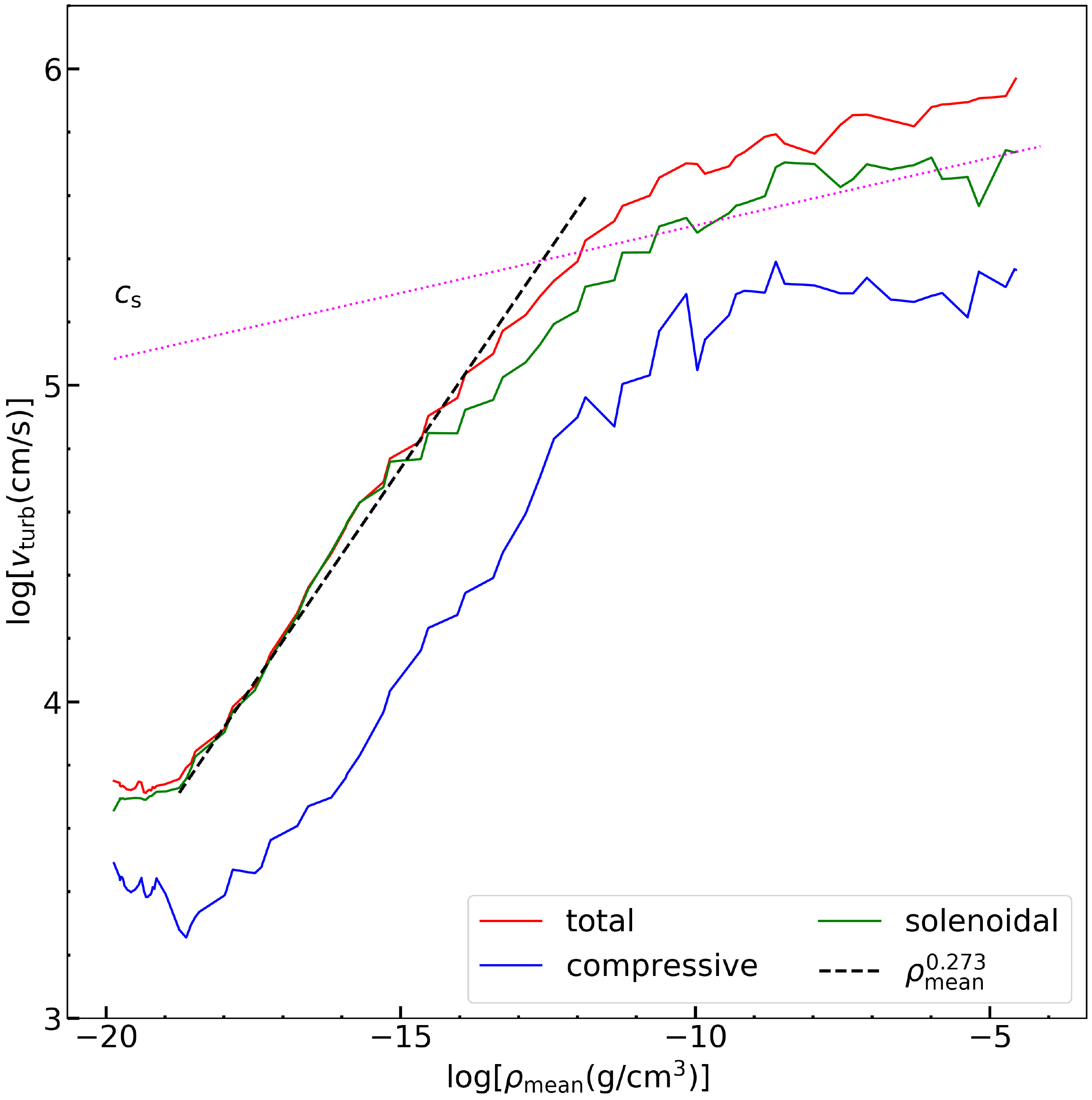}
  \caption{
    Evolution of each modes of the turbulent velocity as a function of the mean density for $\mathcal{M}_0=0.05$ and $\gamma_{\rm eff}=1.09$.
    Red, green, and blue solid {lines correspond} to the total, solenoidal, compressive modes of the turbulent velocity.
    }
  \label{fig:m005mode} 
\end{figure}

As discussed in the previous paragraph, mode coupling between the solenoidal mode and the compressive mode plays important roles.
To show the effects of the mode coupling during the turbulence amplification in detail, we plot each mode of turbulence for $\mathcal{M}_0=0.05$ as a functions of density (Figure \ref{fig:m005mode}).
According to Equation (\ref{eq:vel_al}), the compressive mode of the turbulent velocity is expected to grow slower than the solenoidal mode.

However, we can observe that the compressive mode of the turbulent velocity grows at a similar growth rate to that of the solenoidal mode, although it does not increase in the very early phase. Considering the fact that the growth rate of the compressive mode in the linear analysis is smaller than that of the solenoidal mode (Equation \ref{eq:vel_al}), it is reasonable to regard that the energy conversion from the solenoidal mode to the compressive mode does occur. This should result from the nonlinear term of the equation of motion of the hydrodynamics. As a result, the amplitude of the compressive mode is a few times smaller than that of the solenoidal mode in this calculation throughout the collapse.

%
\subsection{Density fluctuations}\label{sec:density_fl}
\begin{figure}[htbp]
  \centering
  \plotone{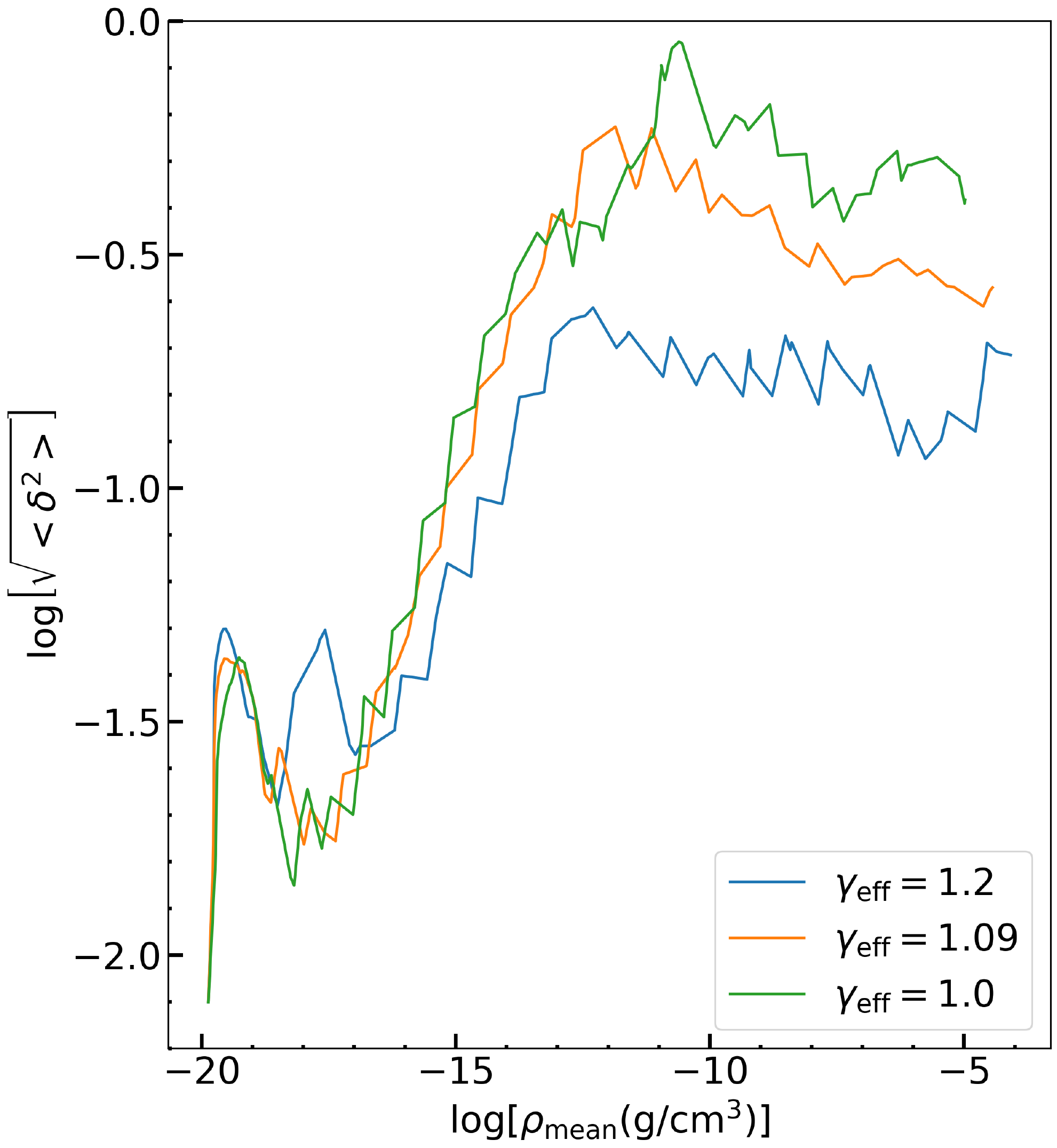}
  \caption{
    Evolution of density fluctuations as a function of mean density for $\mathcal{M}_0=0.1$. 
    {The different colors correspond to different $\gamma_{\rm eff}$.}
    }
  \label{fig:delta-t} 
\end{figure}

In order to understand more about the compressive mode, we investigate density fluctuations in the core,  which is directly related to the compressive turbulence. We plot density fluctuations (Figure \ref{fig:delta-t}) as a function of density $\rho_{\rm mean}$ for $\mathcal{M}_0=0.1$. 
Here the density fluctuation is defined as
\begin{equation}
    \langle \delta^2 \rangle \equiv \sum_{\leq L_{\rm J}/2 } \frac{V_i}{V_{\rm J}}\left(\frac{\rho_i-\rho_{{\rm rad},i}}{\rho_{{\rm rad},i}}\right)^2, 
\end{equation}
where $V_i$ is the volume of the $i$'th cell, and $V_{\rm J}$ is the Jeans volume. $\rho_i$ is the density of the $i$'th cell,  $\rho_{{\rm rad},i}$ denotes the density averaged in each radial shell where $i$'th cell is included. We set the number of radial bins $N_{\rm rad}$ to be 16 in the Jeans Volume as we did for turbulent velocity.\\
We can observe the oscillations of root-mean-square density fluctuations in the core. This oscillation may be due to the overstable behavior of the compressive modes of turbulence.\\
We also notice that the density fluctuations {in all $\gamma_{\rm eff}$} grow initially, but the growth is stopped around $\rho_{\rm mean}\sim 10^{-12}{\rm g/cm^3}$, where the saturation of turbulence occurs (Figure~\ref{fig:vel_p}). As a result, density fluctuations do not grow to the level of $\gg 1$, although {they vary $\sim 0.1 - 0.9$} after the saturation {depending on  $\gamma_{\rm eff}$}.
{These} behavior could be roughly explained in analytic way. As discussed in the previous paragraph, the compressive mode grows at the rate of the solenoidal mode due to the mode coupling. This means $\dot{\delta} \sim \omega \propto a^{-2}$ in the light of Equation (\ref{eq:vel_al}). Consequently, we have $\delta \propto (-t)^{-4/3}$, thus $\delta \propto (-t)^{-1/3}\propto \rho^{1/6}$.
This is the reason for the initial gradual growth of the density fluctuation. After the saturation, growth rate of the vorticity changes as $\omega \propto \rho^{1/2}$. This results in a relation $\delta \propto \ln{(-t)}$, thus the growth is stopped after saturation.

%
%
\subsection{Noise induced by the Discretization}\label{ssec:resolution}
Here, we describe the error introduced by the discretization. In order to understand the nature of the numerical errors, we perform resolution studies.
\begin{figure*}[hbtp]
  \centering
  \plotone{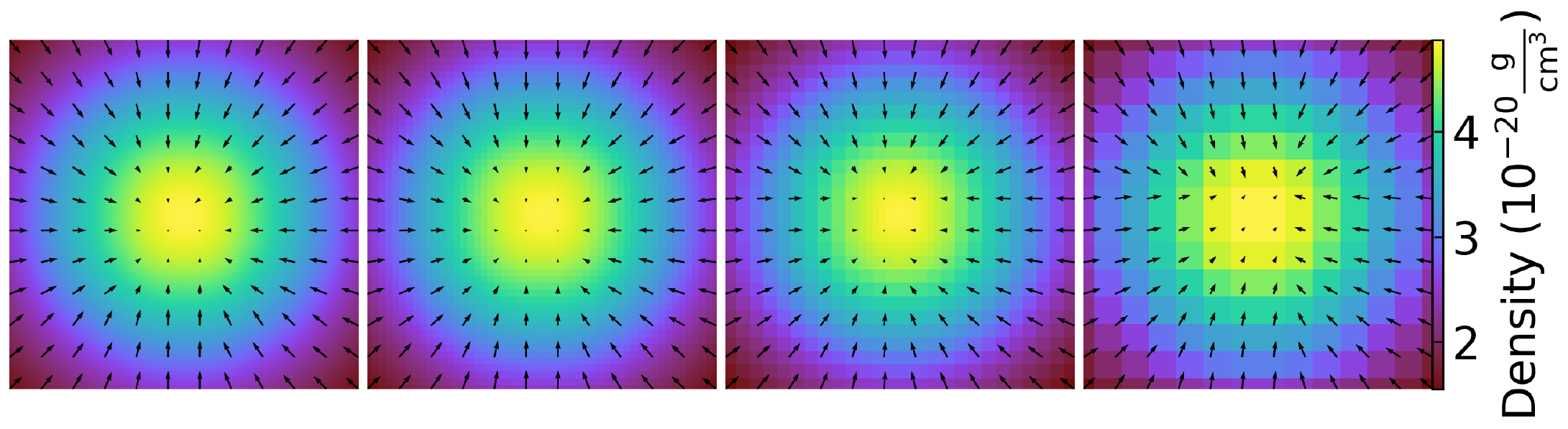}
  \caption{
    Slice plots of density in z=0 at $t=100 ~ {\rm kyr}$ {corresponding to $\rho_{\rm mean} \sim 1.8 \times 10^{-20} ~ {\rm g~cm^{-3}}$} in models RM256 , RM128, RM64 and  RM32. The color legend of density is shown on the right.
    The black arrows are velocity vector, and its length is in arbitrary unit. The box size is $1.0 ~{\rm pc}$. }
  \label{fig:vector} 
\end{figure*}
Figure \ref{fig:vector} shows the velocity field (arrows) overplotted on the density slice (color scales) of  the x-y plane containing the domain center at $t=100 ~ {\rm kyr}$ {corresponding to $\rho_{\rm mean} \sim 1.8 \times 10^{-20} ~ {\rm g~cm^{-3}}$}. Four panels show the results  of  RM256--RM32 models without the initial turbulence ($\mathcal{M}_0 = 0$).
These snapshots correspond to a very early phase of the collapse when the central density increases only by $\sim 5\%$ from the initial density.
The gas clouds will collapse in a spherically symmetric manner if we have an infinite resolution, and the velocity should have only a radial component without any initial turbulence.
However, we can see that non-radial components of velocity fields exist, and RM32 has larger noises than RM256. 
This stems from the Cartesian structure of the grid, and this disturbance is the seeds of  velocity fluctuations in this $\mathcal{M}_0=0$ models.

\begin{figure}[htbp]
  \centering
  \includegraphics[scale=0.3]{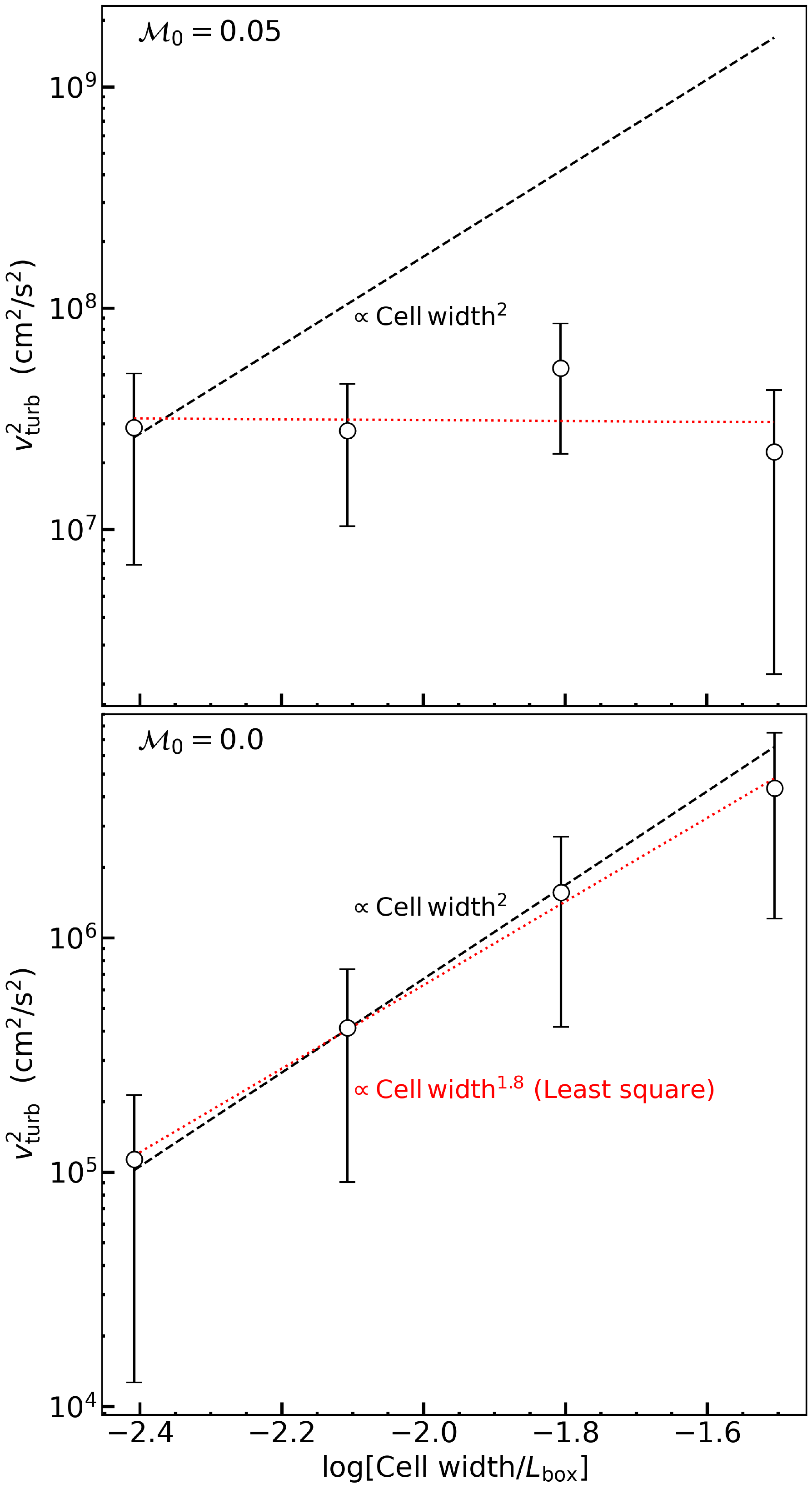}
  \caption{
    Scatter plot for the square of the disturbance vs. non dimensional cell width at $t=100 ~ {\rm kyr}$ {corresponding to $\rho_{\rm mean} \sim 1.8 \times 10^{-20} ~ {\rm g~cm^{-3}}$} in models RM32 - RM256 for $\mathcal{M}_0=0.05$ (upper panel) and for $\mathcal{M}_0=0.0$ (lower panel). A black dashed line corresponds to $\propto ({\rm cell ~ width})^2$.
    Red dotted lines are obtained by a least-square method.} 
  \label{fig:disturbance} 
\end{figure}

Figure \ref{fig:disturbance} shows $v_{\rm turb}^2$ at $t=100 ~ {\rm kyr}$ {corresponding to $\rho_{\rm mean} \sim 1.8 \times 10^{-20} ~ {\rm g~cm^{-3}}$} as a function of the cell size normalized by the box size.

We plot $v_{\rm turb}^2$ with open circles for each resolution model for $\mathcal{M}_0=0.05$ (upper panel) and $\mathcal{M}_0=0.0$ (lower panel) to measure the amplitude of the numerical noise. 
In order to assess the dependence of the noise on the resolution, we obtain the power law fitting function (red dotted lines) with a least-square method.

As a result, for $\mathcal{M}_0=0.05$ there is no resolution dependence so that the turbulence of these models is not affected by the numerical error, while we find the fitted function is proportional to $({\rm cell ~ width})^{1.8}$ for $\mathcal{M}_0 = 0.0$. 
The dependence on the cell width for $\mathcal{M}_0 = 0.0$ means that the error is the second order.  
In absence of the initial turbulence, the specific energy of the turbulence $(\sim v_{\rm turb}^2)$ at the very early stage could be introduced by the transformation of the aspherically symmetric component of the gravitational potential. In fact, \texttt{Enzo}'s gravitational potential solver is a second-order accurate \citep{James77,Enzo}.
Thus, it is reasonable that $v_{\rm turb}^2$ shows a resolution dependence that is proportional to almost the square of the cell width.

\begin{figure*}[htbp]
  \centering
  \includegraphics[scale=0.28]{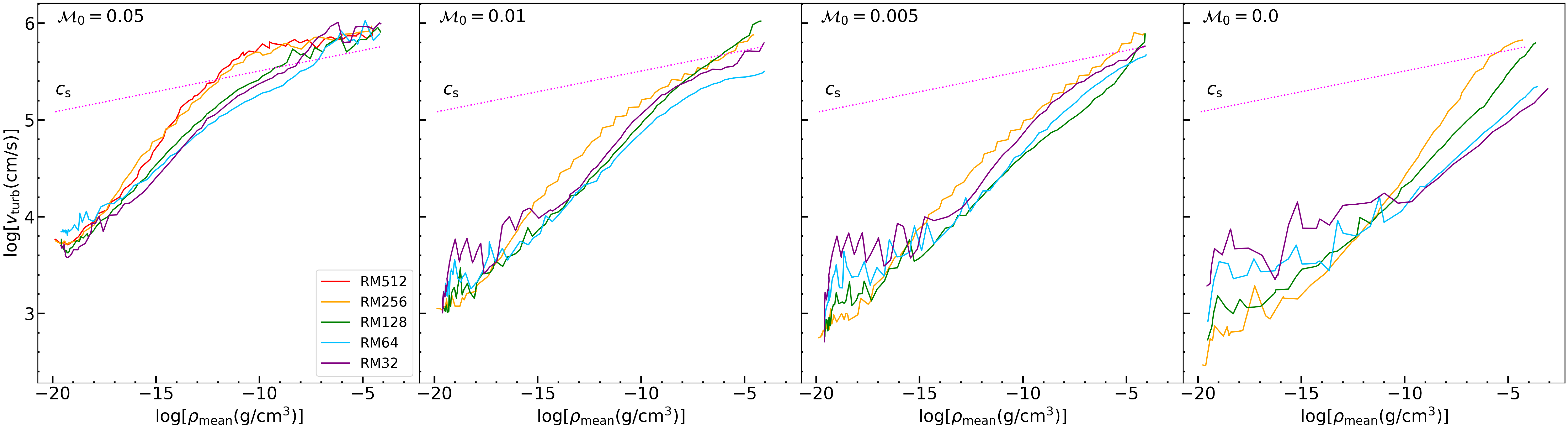}
  \caption{
    Evolution of the turbulent velocity as a function of mean density in models RM32--RM512 for $\gamma_{\rm eff} = 1.09$. RM512 is prepared only for $\mathcal{M}_0=0.05$.}
  \label{fig:reso} 
\end{figure*}

Figure \ref{fig:reso} shows the turbulent velocity for $\gamma_{\rm eff}=1.09$ with different base grid sizes and Jeans Parameters (RM512--RM32). 
The degree of initial turbulence decreases from the left panel to the right in this figure.

In all models, the turbulence is amplified through gravitational contraction.

For $\mathcal{M}_0=0.0$, the initial turbulent velocities are totally numerical error, while the results for $\mathcal{M}_0=0.05$ are physical, as shown in Figure \ref{fig:disturbance}.
Therefore, in Figure 14, we can see whether the results are physical or numerical by comparing the results of the other panels with the values in the rightmost panel.
In the models RM32, RM64, and RM128, the results for $\mathcal{M}_0=0.01$ and $\mathcal{M}_0=0.005$ seem to be numerical because the turbulent velocity around the initial states of these models are comparable to that of $\mathcal{M}_0 = 0.0$. In the model RM256, it is a bit higher than that of $\mathcal{M}_0=0.0$.
Hence, the results of this model may be affected by numerical disturbance, but not fully numerical.
These results suggest the general trend that the resolution required to obtain the physical turbulent velocity increases as the initial Mach number decreases.

As for the $\mathcal{M}_0=0.05$ model, we observe the convergence of RM256 and RM512 in the leftmost panel including the saturated regime. Thus, the obtained results on the growth and the saturation of the turbulent velocity for $\mathcal{M}_0=0.05$ are real.
Considering that the initial turbulent Mach number in a primordial star formation simulation is normally larger than $\mathcal{M}_0 = 0.05$ (e,g,. \citealt{Clark11,Riaz18,katharina20}),
RM256 is enough to resolve the turbulence in those collapse simulations.

%
%

\subsection{Effects of initial spectrum/solenoidal ratio}\label{ssec:test}

We finally test the effects of initial spectrum/solenoidal ratio of turbulence.

We have two sets of simulations here for $\mathcal{M}_0=0.05$ and $\gamma_{\rm eff}=1.09$.
\begin{enumerate}
    \item Test for the different initial turbulent kinetic energy spectrum, such as $E(k) \propto k^{-1}, ~ k^{-2}, ~ k^{-3}$, and $k^{-4}$. The solenoidal ratio is fixed to the natural mixture (Test1).
    \item Test for the different mode mixture for $E(k) \propto k^{-2}$, such as natural mixture, fully solenoidal mode, and fully compressive mode (Test2).
\end{enumerate}

\begin{figure}[htbp]
  \centering
  \plotone{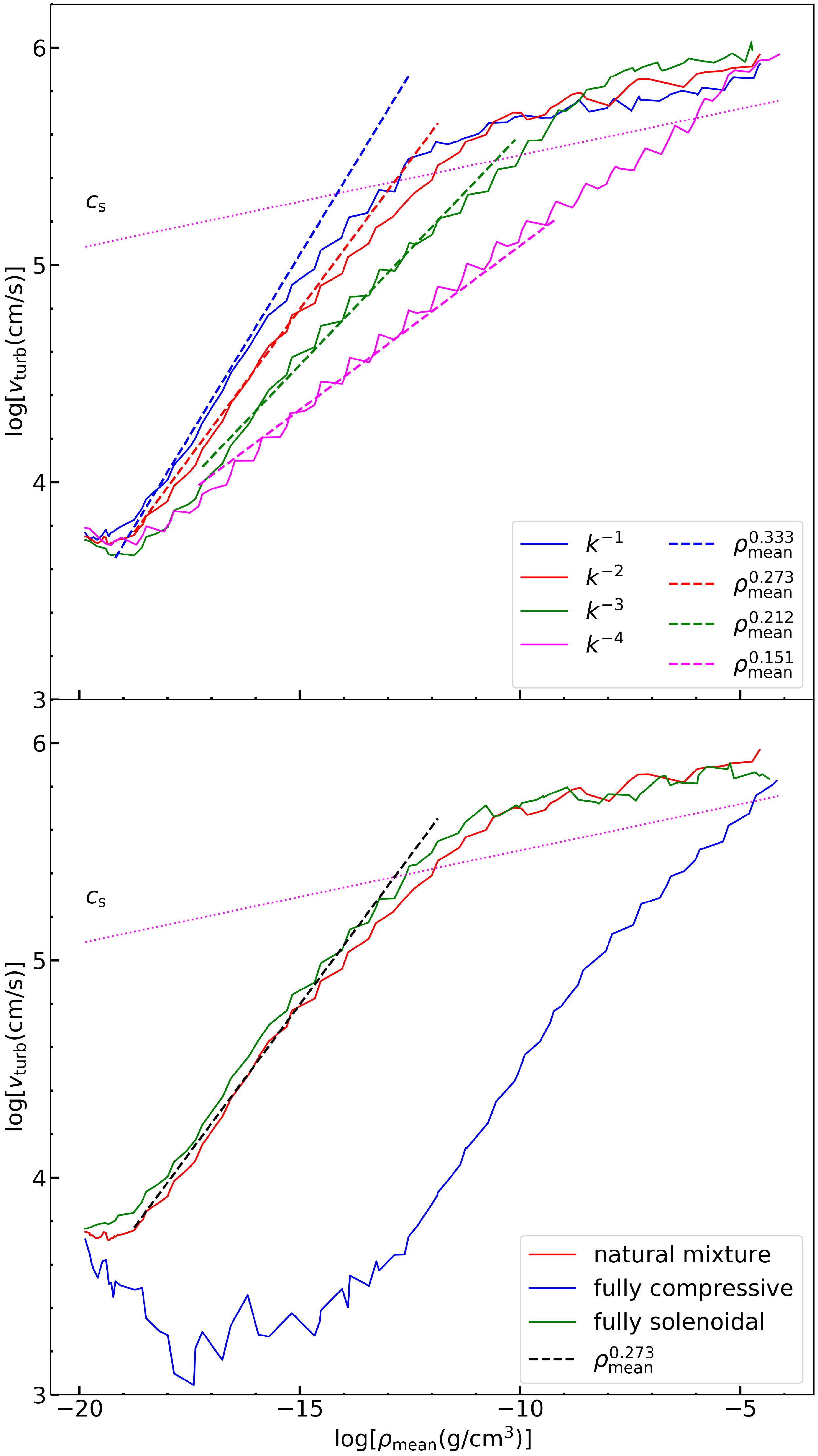}
  \caption{
    Evolution of the turbulent velocity as a function of mean density for Test1 (upper panel) and Test2 (lower panel).
    The solid curves with different colors denote the numerical results for different initial turbulence models.
    The magenta dotted line denotes the average sound speed in the Jeans volume.
    The dashed lines corresponding to each color in the upper panel and the black dashed line in the lower panel depict our analytic estimates (Equation \ref{eq:vel_eff}).}
  \label{fig:mode} 
\end{figure}

We plot the results of these test in Figure \ref{fig:mode}.
This figure shows the evolution of the turbulent velocity as a function of mean density for Test1 (upper panel) and Test2 (lower panel).

\begin{figure}[htbp]
  \centering
  \includegraphics[scale=0.6]{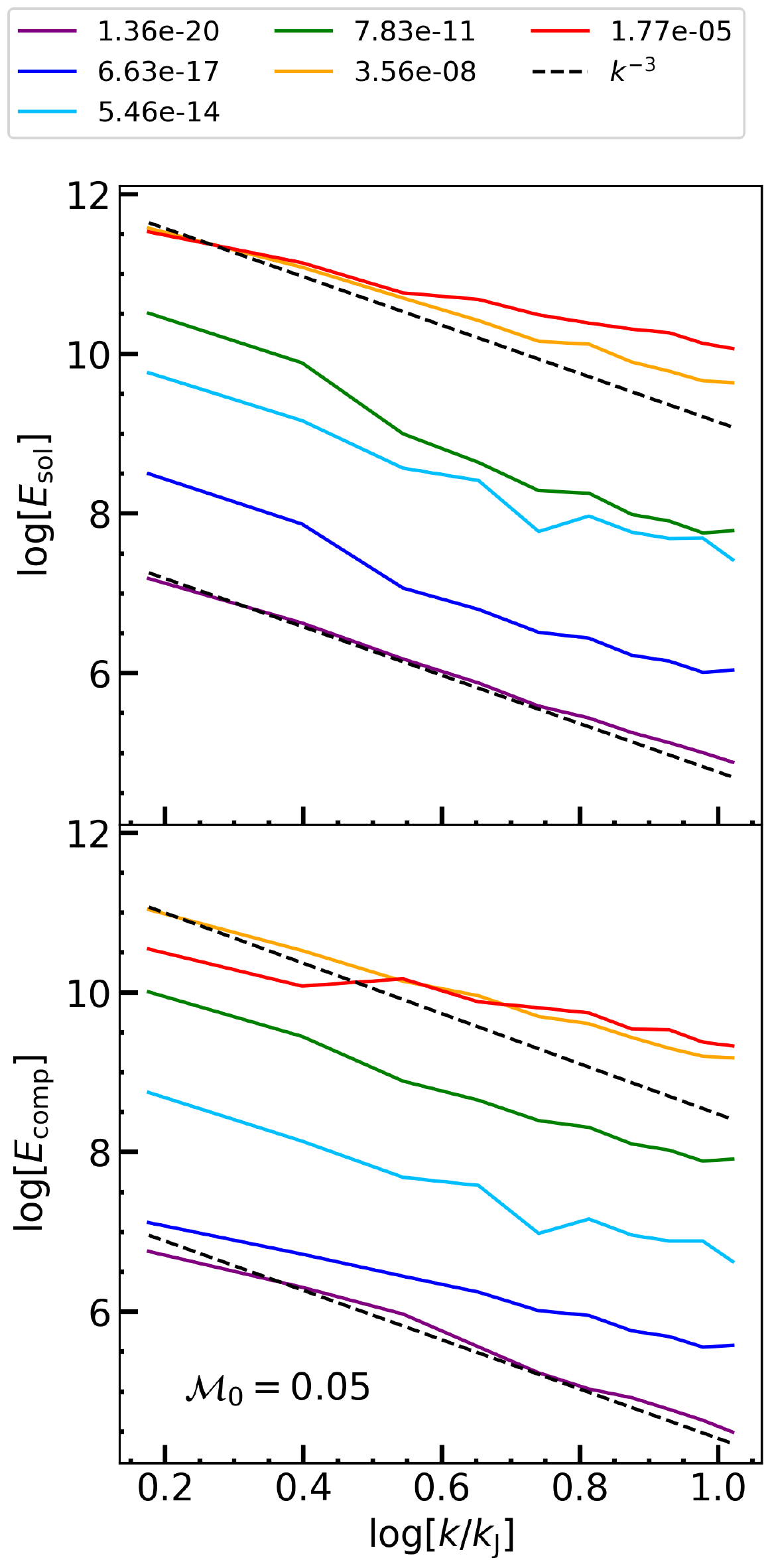}
  \caption{
    Evolution of the turbulent kinetic energy spectra of solenoidal modes (upper panel) and compressive modes (lower panel) for $\mathcal{M}_0=0.05$, respectively.
    The solid curves show the numerical results.
    The black dashed curves denotes $\propto k^{-3}$. The different colors indicate different values of $\rho_{\rm mean}$.
    }
  \label{fig:k3} 
\end{figure}

In the upper panel, we can see that our analytic estimates are in good agreement with all of initial turbulent energy spectra.  This means that the spectral indices do not change throughout the collapse at least on around the Jeans scale $\sim k_{\rm J}$. This conservation of the spectral indices
before the saturation is simply understood by the time scale argument. The time scale of cascade around the core scale ( i.e. Jeans scale ) is $\sim 1/(k_J v_{\rm turb})$, which is longer than the sound crossing time $\sim 1/(k_J c_{\rm s})$ as long as $v_{\rm turb} < c_{\rm s}$. This sound crossing time of the core should be comparable to the gravitational collapse time scale. As a result, the time scale of cascade is longer than the collapse time scale. Thus it is reasonable that the shape of the spectrum is conserved before the saturation {(see also Figure \ref{fig:k3} in details).}

In the lower panel, the natural mixture model and the fully solenoidal model show almost identical results.
The turbulent velocities in these two models are amplified immediately after the collapse begins.
In both cases the turbulent Mach numbers saturate at $\mathcal{M} \sim 2$, being independent on the initial mode mixture of turbulence.

In contrast, the fully compressive model shows a significant difference from above two models.
The turbulent velocity in the fully compressive model decreases immediately after the collapse begins.
This may be due to the oscillation of the small $k$-modes of compressive mode.
The initial density fluctuation is set to be zero, while the (compressive) turbulent velocity is finite.
This means that the amplitude of the initial turbulent velocity is at the local maximum of the oscillation, thereby it decreases initially.
As the collapse proceeds, some of the energy of the compressive mode is converted to the solenoidal mode, which is already seen in Figure \ref{fig:spectrum}. Then the growth of the solenoidal mode is launched, eventually overtaking the compressive mode.  Consequently the turbulent velocity increases with a growth rate similar to the other two models, i.e. the growth rate of solenoidal mode.
This means that the growth of the solenoidal mode finally dominates that of the compressive mode during the contraction even in the absence of the solenoidal mode at the onset of the collapse.  

{To reinforce the above argument for the upper panel in Figure \ref{fig:mode},
we also plot the evolution of the turbulent kinetic energy spectra (Figure \ref{fig:k3}) for the model of initial energy spectrum $E(k) \propto k^{-3}$ in Test1.
The details of this figure are the same as Figure \ref{fig:spectrum} except for the initial turbulent kinetic energy spectrum.
In the upper panel, we can clearly see the conservation of spectral indices before the saturation ($\rho_{\rm mean} < 10^{-8}$).
In contrast, after the saturation, the spectral indices gradually increase (i.e. become shallower power-law distribution), starting from the high-$k$ end.
It is because smaller turbulent eddies at the high-$k$ end firstly redistribute the energy in the $k$-space by the nonlinear effect, since the eddy time scales are shorter than those in low-$k$ eddies.
}

Summarizing the results of these test, we find that the growth rate of the turbulent velocity sensitively depends on the initial power-law index of the turbulent kinetic energy spectrum.
On the other hand, the growth rate is insensitive to the initial solenoidal ratio, following the growth rate of the solenoidal mode (Equation 14), although the launch of the growth is delayed for initially 100\% compressive turbulence model.
In all cases, the turbulent velocities grow, in agreement with the analytic estimate in Section \ref{sec:estimate}.

\section{Discussion} \label{sec:discussion}
We have set weak initial turbulence ($\mathcal{M}_0 \leq 0.1$) to understand the detailed evolution of turbulence.
Cosmological simulations of other authors have shown that the gas inflow along the filaments of the halo already generates turbulence of $\mathcal{M}_0\gtrsim 0.5$ in the low-density phase ($n_{\rm H} \sim 10^3 ~ {\rm cm^{-3}}$) before the run-away collapse begins (e.g., \citealt{Wise07,Greif08,Greif11,Greif2012}).  These initial Mach numbers clearly exceed the threshold found in this paper (Equation \ref{eq:M0cr}), thereby the star forming dense core should reach the sonic point during the  collapse. This is consistent with the previous numerical studies such as \citet{Greif2012}.  In addition, the saturation level of the turbulent velocities for $\gamma_{\rm eff}=1.09$ seems to be supersonic (Figure \ref{fig:vel_p}), that is also consistent with the results of \citet{Greif2012}, in which the Mach number at the final phase is $\mathcal{M} \simeq 2$ .  Hence, our finding in this paper can well describe the numerical studies on first star formation so far.

We have used the initial conditions with an ideal Bonnor-Ebert sphere, which neglects the initial rotation and the asymmetry of clouds. Consequently, we have minimized the effects of the deformation and the shear motion triggered by the rotation. We also assume constant $\gamma _{\rm eff}$,  to simplify the entropy production/reduction processes, such as hydrodynamical shocks or radiative cooling.  The presence of shock waves leads to the time variation of vorticity due to baroclinic term in the vorticity equation, which is neglected in this work.
These assumptions could have some impact on the present results, which can be discussed by  the comparison with a previous study.
\cite{Greif2012} followed the gas collapse until protostellar cores form from cosmological initial conditions, considering various cooling processes and shock heating.
Their simulations also result in the amplification of turbulence up to $\mathcal{M} \sim 2$, which is consistent with the results in this paper (Figure \ref{fig:vel_p}) although they only consider the cases starting  from transonic turbulent velocities.  This implies that the initial gas rotation and entropy variation have little  impacts on the amplification of turbulence.
{Additionally, \cite{Turk2012} also followed the gas collapse from cosmological initial conditions as well as \cite{Greif2012} except for the presence of  magnetic fields and the difference of a dynamic range.
As a result, the density vs. vorticity squared relation in their calculations almost follows $\omega^2 \propto \rho^{4/3}$. This fact shows that the contribution of the baroclinic term in the vorticity equation is small in the collapse simulations.}

In this paper, we have not investigated the turbulence in the mass accretion phase, which is important for the dynamics of the disk fragmentation.
The disks would form around hydrostatic cores (protostars) where radiative cooling is inefficient in the optically thick regime with densities $\rho_{\rm peak} \gtrsim 10^{-5} ~ {\rm g~cm^{-3}}$ \citep{Larson69,Penston69}.
Recent studies have shown that the initial turbulence can enhance disk fragmentation, and can affect the initial mass function (IMF) of the first generation of stars (e.g., \citealt{Clark11,Riaz18,katharina20}).
Because of the strong shear motion in the disk and the chaotic motions caused by the fragments, it is reasonable to expect that the  turbulence is kept driven in the mass accretion phase.



Turbulent motion should be accompanied by magnetic fields, while it is not taken into account in the present paper. Presence of the magnetic fields is important for the protostellar evolution in general. It can introduce additional heating during the run-away collapse phase \citep{schleicher09,nakauchi19,Nakauchi21}, and magnetic breaking/outflow launching is activated in the mass accretion phase \citep{Machida13}.
High-resolution numerical simulations show that, in the turbulent primordial gas with magnetic fields, the small-scale dynamo effect can amplify the seed magnetic fields to certain levels depending on the initial seed field strength and on the numerical resolution \citep{Sur10,Sur12,Fed11,Turk2012}.
However, the numerical results do not converge even with the state-of-art highest-resolution simulations, because the eddy scale is too small compared to the Jeans scale to be resolved numerically.
Therefore, the actual magnitude of the magnetic fields during the collapse phase in the first star-forming environment is still unknown. One possible direction to overcome this numerical difficulty is to introduce a sub-grid model based upon the analytic estimates \citep{Schleicher2010,Schober2012,Xu16,Xu20,Mackee20}.


 Finally, it is worthy to note that the present results are quite general and robust, thereby can be applied to not only the primordial case, but also to more general star formation/cloud collapse processes. Normally, cloud collapse simulations do not use large Jeans parameters, so that they fail to resolve the turbulent eddies, which could change the physical process in the core. Hence we have to keep in mind that run-away collapsing core tend to be turbulent, even if the initial seed velocity field is small. 
 

\section{Summary} \label{sec:summary}
We study the amplification of turbulence in collapsing gas clouds by performing high-resolution numerical simulations until the gas peak density reaches $\rho_{\rm peak} = 10^{-4} ~ {\rm g~cm^{-3}}$.
We find that the turbulence can be amplified through the contraction of the gas cloud.
We analytically estimate the scaling relation between the turbulent velocity/vorticity and  density, and find that our simulation results are in good agreement with our analytic estimates, although special care is necessary on the introduced error by grid discretization.
We also find a critical initial Mach number at $\rho_0$ to achieve sonic/super sonic turbulence at a given density $\rho_{\rm sonic}$.  For  $\rho_0=10^{-19}~{\rm g~cm^{-3}}$ and  $\rho_{\rm sonic}=10^{-6}~{\rm g~cm^{-3}}$, we obtain $\mathcal{M}_{\rm 0, cr}\simeq 0.001$. As a result, highly turbulent dense cores are logically expected, since this initial Mach number is easily realized in cosmological simulations.   
To investigate the amplification mechanism in more detail, we follow the evolution of solenoidal and compressive velocity modes separately.
The solenoidal mode is dominant from just after we start the simulations if we subtract the radial velocity from the total velocity. 
This indicates that the gravitational compression of clouds powers the amplification of solenoidal modes, and the total turbulent velocity increases.
The solenoidal modes continue to grow until we terminate the simulations, and
the turbulent velocity eventually reaches the supersonic velocity even with small initial Mach number $\mathcal{M}_0 =0.05$.
We finally test the effects of the initial spectrum/solenoidal ratio of the turbulence.
It is found that the growth rate of the turbulence sensitively depends on the initial spectrum of the turbulence, which is also in agreement with our analytic estimate. On the other hand, the growth rate is same among the various initial solenoidal ratio cases, simply because the growth rate is basically dominated by the solenoidal mode.
These results indicate that 
the turbulence can play important roles in collapsing clouds in general cases.
\acknowledgments
We are grateful to anonymous referee for careful reading of the manuscript and constructive comments.
We thank K.Shima, K.Tomida, N.Yoshida, T.Matsumoto, and J.H.Wise for fruitful discussions and useful comments. We are thankful for the support by Ministry of Education, Science, Sports and Culture, Grants-in-Aid for Scientific Research No. 17H02869, 17H01101, and 17H06360.
GC is supported by Overseas Research Fellowships of the Japan Society for the Promotion of Science (JSPS) for Young Scientists.
A part of numerical calculations in this work was carried out on Yukawa-21 at the Yukawa Institute Computer Facility.
Computations and analysis described in this work were performed using the publicly-available \texttt{Enzo} and \texttt{yt} codes, which is the product of a collaborative effort of many independent scientists from numerous institutions around the world.



\bibliographystyle{aasjournal}
\vspace{5mm}

\listofchanges
\end{document}